\documentclass[]{jfm}
\usepackage{graphicx}

\usepackage[latin1]{inputenc}
\usepackage{amsmath}
\usepackage{amsfonts}
\usepackage{amssymb}
\usepackage{makeidx}
\usepackage{graphicx}
\usepackage{tikz}
\usepackage{floatrow}
\usepackage[export]{adjustbox}
\usepackage{booktabs}
\usepackage{multirow}
\usepackage[percent]{overpic}
\usepackage{subfig}

\shorttitle{Near-onset dynamics in natural doubly diffusive convection}
\shortauthor{C. Beaume, A. M. Rucklidge and J. Tumelty}

\title{Near-onset dynamics in natural doubly diffusive convection}

\author{C\'edric Beaume\aff{1},
	Alastair M. Rucklidge\aff{1}
	\and Joanna Tumelty\aff{1}}

\affiliation{\aff{1}School of Mathematics, University of Leeds, Leeds LS2 9JT, UK}

\begin{document}
	
	\maketitle
	
	\begin{abstract}
		Doubly diffusive convection is considered in a vertical slot where horizontal temperature and solutal variations provide competing effects to the fluid density while allowing the existence of a conduction state.
		In this configuration, the linear stability of the conductive state is known, but the convection patterns arising from the primary instability have only been studied for specific parameter values. 
		We have extended this by determining the nature of the primary bifurcation for all values of the Lewis and Prandtl numbers using a weakly nonlinear analysis.
		The resulting convection branches are extended using numerical continuation and we find large-amplitude steady convection states can coexist with the stable conduction state for sub- and supercritical primary bifurcations.
		The stability of the convection states is investigated and attracting travelling waves and periodic orbits are identified using time-stepping when these steady states are unstable. 
	\end{abstract}
	
	\begin{keywords}
		Authors should not enter keywords on the manuscript, as these must be chosen by the author during the online submission process and will then be added during the typesetting process (see http://journals.cambridge.org/data/\linebreak[3]relatedlink/jfm-\linebreak[3]keywords.pdf for the full list)
	\end{keywords}
	
	
	\section{Introduction}
	Doubly diffusive convection can occur when a binary fluid is subject to external gradients of temperature and of concentration.
	It has primarily been studied in the context of oceanography as an important mechanism for heat and salt transport \citep{huppert1981,schmitt1994double}, since approximately $44\%$ of the world's oceans are known to display this phenomenon \citep{you2002}.
	Doubly diffusive convection can display a wealth of behaviour that depends on the respective orientations of the (thermal and solutal) driving gradients. 
	At low latitude, oceans typically feature thermohaline staircases \citep{schmitt1987csalt, schmitt2005enhanced}, where the flow is characterised by well-mixed horizontal layers interspersed with interfaces displaying sharp upward pointing gradients of temperature and salinity.
	In configurations forced by upward gradients of salinity and temperature, fluids display strikingly complex dynamics characterised by an alternation of well-mixed convection zones and fingers transporting salt mostly vertically \citep{Krishnamurti2003,Krishnamurti2009}.
          This salt fingering instability is a natural mechanism that enhances the local mixing of the oceans \citep{schmitt1994double}.
	At high latitude, the forcing gradients typically point downwards and perturbations to the thermohaline staircase give rise to oscillatory dynamics in a behaviour called diffusive layering \citep{kelley2003,santos2014}.
	Similar doubly diffusive phenomena are also found at the Earth's core-mantle boundary \citep{Lay2008}, in astrophysical flows \citep{spiegel1969,spiegel1972,bethe1990} and in processes involving solidification \citep{wilcox1993}, such as in magma crystallisation \citep{huppert1984}. 
	
	Originally motivated by the above configurations, doubly diffusive convection has become a paradigm for the study of fluids as dynamical systems.
	A large variety of flow states comprised of convection rolls have been identified, including standing, travelling and modulated waves \citep{deane1988,kolodner1991,Predtechensky1994}.
	Temporal complexity has also been found in various forms \citep{spina1998,batiste2001} and is generated in a number of ways \citep{knobloch1986,rucklidge1992,beaume2020transition}.
	Work focusing on steady state dynamics also revealed intricate phenomena like spatial localisation in the presence \citep{mercader2009,mercader2011} and in absence \citep{beaume2011} of Soret effect.
	Localised convection states, called convectons, are found on solution branches exhibiting oscillatory trajectories in parameter space in a behaviour called snaking \citep{knobloch2015}.
	Travelling versions of convectons have also been found and produce an interesting hierarchy of interconnected instabilities \citep{watanabe2012,watanabe2016}.
	
	Practical considerations led to the study of inclined domains, where gravity and the driving gradients are no longer aligned \citep{paliwal1980a,paliwal1980b,bergeon1999}, as well as cases in which the salinity and temperature gradients are not aligned with each other \citep{tsitverblit1993multiplicity,tsitverblit1995bifurcation,dijkstra1996bifurcation}.
	Motivated by solidification fronts \citep{wilcox1993} and mixing currents in the vicinity of icebergs \citep{huppert1981}, this article focuses on a configuration, typically referred to as {\it natural doubly diffusive convection}, where the driving gradients are aligned but orthogonal to gravity. 
	
	The bifurcation scenario for a range of small aspect-ratio domains was elucidated by \citet{xin1998bifurcation} and \citet{bergeon2002natural}, and large aspect-ratio domains were found to support the existence of spatially localised states \citep{bergeon2008spatially}.
	More recent work focused on a full characterisation of these localised states and on the emergence of chaos in large aspect-ratio domains \citep{beaume2013convectons,beaume2013b,beaume2018three,beaume2020transition}.
	Most of the pattern formation introduced in the above references can be found close to onset but they have, mostly, been studied for a certain set of parameter values.
        Despite the fact that a comprehensive linear stability analysis in the special case of balancing thermal and solutal gradients has been available for more than two decades \citep{ghorayeb1997double}, the analysis for unbalanced gradients was only recently attempted by \citet{shankar2021stability}.
	Further, little is known about the dynamics near onset in the balanced case besides its linear regime, which makes it difficult to extrapolate the dynamics of the system away from the parameter values used in previous studies.
	  Here, we perform a weakly nonlinear analysis and augment it by numerically continuing branches of spatially periodic states in a small-aspect ratio domain.
          Our present work extends the previous analyses to a wider range of parameter values. 
	In the next section, we present the mathematical framework associated with our case of doubly diffusive convection.
	In Section 3, we detail the weakly nonlinear analysis of this system, followed by a characterisation of the nonlinear dynamics in Section 4.
	We conclude in Section 5 with a short discussion.
	
	\section{Mathematical formulation}\label{sec:mathematical_formulation}
	
\begin{figure}
		\centering
		\includegraphics[]{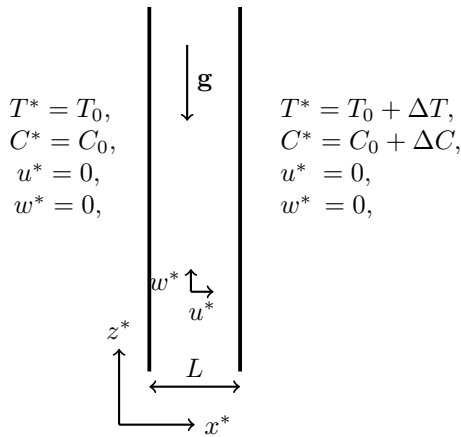}
		\caption{Sketch of the two-dimensional domain of interest together with the dimensional form of the boundary conditions.}
		\label{fig:domain}
	\end{figure}

	We consider the natural doubly diffusive convection of an incompressible fluid in a two-dimensional domain with periodic boundary conditions in the vertical direction.
	The side walls are rigid, impermeable and maintained at fixed temperatures and solutal concentration.
	The right wall is held at a higher temperature ($T_0+\Delta T$) and solutal concentration ($C_0+\Delta C$) than the left wall, where the temperature is $T_0$ and the solutal concentration is $C_0$.
	This configuration is depicted in figure~\ref{fig:domain}. 
	
	The system is governed by the Navier--Stokes equation for fluid momentum, the incompressibility condition and advection-diffusion equations for both the temperature and the concentration.
	Cross-diffusion due to the Soret and Dufour effects is not considered.
	The imposed temperature and solutal concentration differences are assumed to be small enough so that the Boussinesq approximation can be applied, whereby density variations are neglected except in buoyancy terms. 
	The density of the fluid is assumed to have a linear dependence on its temperature and concentration:
	\begin{equation}
	\rho^* = \rho_0 + \rho_T(T^*-T_0) + \rho_C(C^* - C_0),
	\end{equation}
	where $\rho_0$ is the density of the fluid at temperature $T_0$ and concentration $C_0$ and $\rho_T<0$ (resp. $\rho_C>0$) is the thermal (resp. solutal) expansion coefficient.
	
	We introduce the non-dimensional quantities:
	\begin{equation}
	\mathbf{x} = \frac{\mathbf{x}^*}{L},\quad t = \frac{t^*}{L^2/\kappa}, \quad \mathbf{u} = \frac{\mathbf{u}^*}{\kappa/L}, \quad T= \frac{T^*-T_0}{\Delta T}, \quad C = \frac{C^*-C_0}{\Delta C},\quad p = \frac{p^*}{\rho_0 \kappa \nu/L^2}, \label{eq:nondim}
	\end{equation}
	where $L$ is the wall separation, $\kappa$ is the rate of thermal diffusivity and $\nu$ is the kinematic viscosity.
	The non-dimensional governing equations for the fluid velocity $\mathbf{u} = u \mathbf{\hat{x}} + w \mathbf{\hat{z}}$, the pressure $p$, the temperature $T$ and the concentration $C$ thus read:
	\begin{align}
	\dfrac{1}{Pr} \left(\dfrac{\partial \mathbf{u}}{\partial t} + \mathbf{u}\cdot \nabla \mathbf{u}\right) &= -\nabla p +\nabla^2 \mathbf{u} + Ra\left(T+NC\right)\mathbf{\hat{z}} , \label{eq:NS}\\
	\nabla \cdot \mathbf{u} &= 0, \label{eq:incomp}\\
	\dfrac{\partial T}{\partial t} + \mathbf{u}\cdot \nabla T &= \nabla^2 T, \label{eq:T}\\
	\dfrac{\partial C}{\partial t} + \mathbf{u}\cdot \nabla C &= \frac{1}{Le}\nabla^2 C, \label{eq:C}
	\end{align}
	where $\mathbf{\hat{z}}$ is the vertical ascending unit vector and where we have introduced the following dimensionless parameters:
	\begin{eqnarray}
	\textrm{the Prandtl number~}&Pr &= \frac{\nu}{\kappa},\\
	\textrm{the Rayleigh number~}&Ra &= \frac{gL^3|\rho_T|\Delta T}{\rho_0\nu \kappa},\\
	\textrm{the buoyancy ratio~}&N &= \frac{\rho_C \Delta C}{\rho_T \Delta T},\\
	\textrm{and the Lewis number~}&Le &= \frac{\kappa}{D},
	\end{eqnarray}
	where $D$ is the rate of solutal diffusivity. 
	The non-dimensional boundary conditions read:
	\begin{align}
	u = 0,\quad w = 0, \quad -\frac{\partial p}{\partial x} + \frac{\partial^2 u}{\partial x^2}=0, \quad T = 0, \quad C = 0 \quad \text{ on } \quad x = 0, \label{eq:bc0}\\
	u = 0,\quad w = 0,\quad -\frac{\partial p}{\partial x} + \frac{\partial^2 u}{\partial x^2}=0,\quad T = 1, \quad C = 1 \quad \text{ on }\quad x = 1, \label{eq:bc1}
	\end{align}
	where the pressure boundary condition is the projection of the Navier--Stokes equation on the boundary.
	Each variable is periodic in the vertical direction.
	
	We restrict our attention to the case $N = -1$, where the full system (\ref{eq:NS}--\ref{eq:C}, \ref{eq:bc0}, \ref{eq:bc1}) admits the steady conduction state with linear temperature and concentration profiles between the side walls:
	\begin{equation}
	\mathbf{u}=\mathbf{0},\quad T = x,\quad C = x. \label{eq:conduction_state}
	\end{equation}
	We further introduce convective variables as the departures of the temperature and concentration from the conduction state:
	\begin{align}
	\Theta &= T-x,\label{eq:T_dev}\\
	\Phi &= C - x.\label{eq:C_dev}
	\end{align}
	Using these new variables, the conduction state takes the form
	\begin{equation}
	\textbf{u} = \mathbf{0}, \quad \Theta = 0, \quad \Phi = 0,\label{eq:conduction_state_2}
	\end{equation}
	and system (\ref{eq:NS})--(\ref{eq:C}) can be written as:
		\begin{align}
	\dfrac{1}{Pr} \left(\dfrac{\partial \mathbf{u}}{\partial t} + \mathbf{u}\cdot \nabla \mathbf{u}\right) &= -\nabla p +\nabla^2 \mathbf{u} + Ra\left(\Theta-\Phi\right)\mathbf{\hat{z}} , \label{eq:NS_conv}\\
	\nabla \cdot \mathbf{u} &= 0, \label{eq:incomp_conv}\\
	\dfrac{\partial \Theta}{\partial t} + \mathbf{u}\cdot \nabla \Theta &= -u + \nabla^2 \Theta, \label{eq:T_conv}\\
	\dfrac{\partial \Phi}{\partial t} + \mathbf{u}\cdot \nabla \Phi &= - u + \frac{1}{Le}\nabla^2 \Phi. \label{eq:C_conv}
	\end{align}

	This formulation involving the convective variables allows the identification of two symmetries of the system: the reflection $S_{\Delta}$ and the continuous translation $T_{\delta}$:
	\begin{align}
	S_{\Delta}:\quad(x,z)\mapsto(1-x,-z), \quad(u,w,\Theta,\Phi)&\mapsto -(u,w,\Theta,\Phi), \label{eq:symm}\\
	T_{\delta}:\quad(x,z) \mapsto(x,z+\delta), \qquad(u,w,\Theta,\Phi) &\mapsto (u,w,\Theta,\Phi).
	\end{align}
	With periodic boundary conditions, these generate the symmetry group $\mathcal{O}(2)$ and restrict the types of bifurcation that can occur from the conduction state.
	
	\section{Weakly nonlinear predictions}\label{sec:results}
	To predict the pattern formation present in our system, we start by performing the linear stability analysis of the conduction state $(u,w,p,\Theta,\Phi) = (0,0,0,0,0)$, which was previously done by \citet{ghorayeb1997double} and by \citet{xin1998bifurcation}.
	We briefly rederive their results in the following subsection so that they can be applied in the later weakly nonlinear analysis, where we derive Ginzburg\textendash{}Landau equations to model the small-amplitude behaviour close to the primary bifurcation for all Lewis and Prandtl numbers.
	
	\subsection{Linear stability analysis}\label{sec:linstab}
	We first consider small-amplitude stationary normal mode perturbations to the conduction state:
	\begin{equation}
	(u,w,p,\Theta,\Phi)^T = \epsilon\left(\left(U_1(x),W_1(x),P_1(x),\Theta_1(x),\Phi_1(x)\right)^Te^{ikz} + c.c. \right) + O(\epsilon^2), \label{eq:linstab_disturbance}
	\end{equation}
	where  $c.c.$ denotes the complex conjugate of the preceding term, $\epsilon \ll 1$, $k$ is the vertical wavenumber and $\lambda$ is the temporal growth rate of the perturbation.
	Inserting expansion (\ref{eq:linstab_disturbance}) into system (\ref{eq:NS_conv})--(\ref{eq:C_conv}) and linearising the resulting system yields the eigenvalue problem:
	\begin{equation}
	\mathcal{L}(Ra)\mathbf{\Psi_1} = \mathbf{0}, \label{eq:linsystem}
	\end{equation}
	for $Ra$ and $\mathbf{\Psi}_1$ where
	\begin{equation}
	\mathbf{\Psi_1} = \left(U_1,W_1,P_1,\Theta_1,\Phi_1\right)^Te^{ikz} + c.c.,
	\end{equation}
	and
	\begin{equation}
	\mathcal{L}(Ra) = \begin{pmatrix}
	\nabla^2 & 0 & -\partial_x & 0 & 0 \\
	0 & \nabla^2 & -\partial_z & Ra & -Ra \\
	\partial_x & \partial_z & 0 & 0 & 0 \\
	-1 & 0 & 0 & \nabla^2 & 0 \\
	-1 & 0 & 0 & 0 & \frac{1}{Le}\nabla^2
	\end{pmatrix}.\label{eq:def_L}
	\end{equation}
	The complex functions $U_1$, $W_1$, $P_1$, $\Theta_1$ and $\Phi_1$ satisfy Dirichlet boundary conditions on the side walls for the velocity, temperature and concentration perturbations:
	\begin{equation}
	U_1(x) = W_1(x) = \Theta_1(x) = \Phi_1(x) = 0\quad \text{ on } x = 0,1,
	\end{equation}
	and the projection of the Navier--Stokes equation onto the boundary for the  pressure perturbation:
	\begin{equation}
	0 = -\frac{\partial P_1}{\partial x} + \frac{\partial^2U_1}{\partial^2x} \quad \text{ on } x = 0,1.\label{eq:p_bc}
	\end{equation}

\begin{figure}
		\centering
		\includegraphics{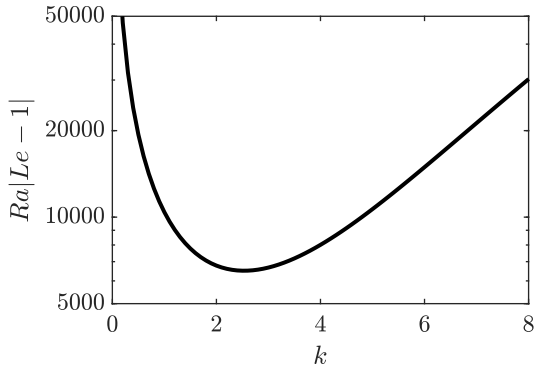}
		\caption{Marginal stability curve for the onset of doubly diffusive convection. The conduction state is stable to modes with wavenumber $k$ below the curve, and unstable to them above. The minimum of this curve is $Ra_c|Le-1| \approx 6509$ with wavenumber $k_c \approx 2.5318$ and corresponds to the location of the primary bifurcation.}
		\label{fig:singlemargstabcurve}
	\end{figure}

\begin{figure}
\centering
\includegraphics[]{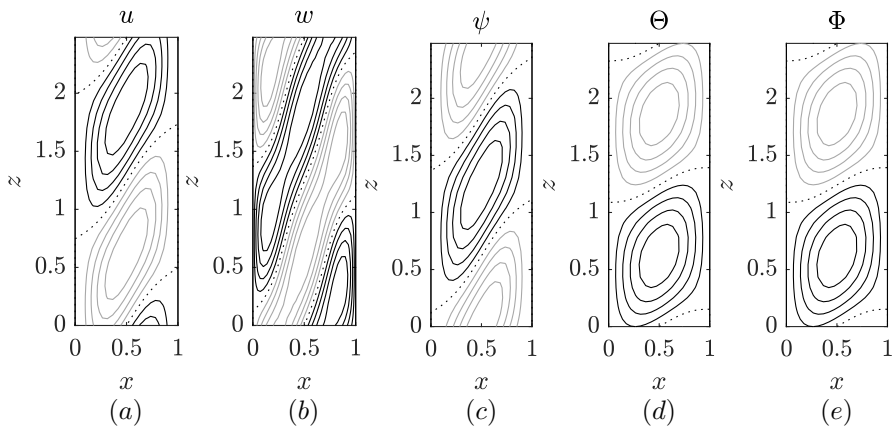}
		\caption{Contour plots of a single wavelength of the real critical eigenvector $\mathbf{\Psi}_1$ for $Le = 11$ ($k_c \approx 2.53$). 
		The profiles show the perturbations in (a) horizontal velocity, $u$, (b) vertical velocity, $w$, (c) velocity streamfunction, $\psi$, where $u = -{\psi}_z$ and $w = {\psi}_x$, and in the convective variables (d) $\Theta$ and (e) $\Phi$.
		Black (grey, dotted) lines indicate positive (negative, zero) values and are separated by 20\% of the maximum absolute value.}
		\label{fig:O(eps)soln}
	\end{figure}

	Solutions to (\ref{eq:linsystem}--\ref{eq:p_bc}) are independent of $Pr$ and satisfy $\Phi_1 = Le\,\Theta_1$.
	Consequently, the only parameter dependence in the linear problem comes from the buoyancy term in the momentum equation, which takes the form $Ra(1-Le)\Theta_1$.
	The accordingly simplified version of equation (\ref{eq:linsystem}) is then solved using a spectral eigenvalue solver based on a Chebyshev--Legendre collocation method for a range of $k$ to determine the marginal stability curve in figure~\ref{fig:singlemargstabcurve}.
	This curve reveals a minimum at $k_c \approx 2.5318$ and $Ra_c|Le-1| \approx 6509$, which corresponds to the primary instability of the conduction state. 
	The absolute value here comes from the fact that the system resulting from left wall heating and that resulting from right wall heating are equivalent.
	Contour plots presenting the fields for the velocity components, streamfunction and convective variables of this eigenmode for $Le=~11$ are shown in figure~\ref{fig:O(eps)soln}.
	The conduction state is thus first unstable to a spatially periodic state constituted of counter-rotating convection rolls that, when $Le > 1$, slant downwards from the hotter wall, filling the domain and extending to the cold wall.
	
	The conduction state can also undergo Hopf bifurcations, where the growth rate is purely imaginary.
	However, for the parameter values tested, these bifurcations occurred for Rayleigh numbers that are orders of magnitude larger than that of the primary stationary bifurcation and are therefore out of the scope of the present work. 
	
	\subsection{Weakly nonlinear analysis}\label{sec:weaklynonlinear}
	To investigate the weakly nonlinear regime around this primary bifurcation, we set $Ra = Ra_c + \epsilon^2 r$ with $r = \mathcal{O}(1)$ and $\epsilon \ll 1$ and assume that the system evolves on a slow temporal scale $T_1 = \epsilon^2 t$. We also introduce a long spatial scale, $Z = \epsilon z$, to allow small-amplitude states with long spatial modulations.
	The small-aspect ratio domains considered in the numerical computations in Section~\ref{sec:nonlinearanalysis} do not permit these long-scale modulations, so terms involving derivatives with respect to $Z$ may be ignored in the subsequent analysis with no effect on the main result of this section: the criticality of the primary bifurcation.
	However, this long spatial variable has been included here to broaden the scope of the analysis and will be considered in future work.
	We emphasise that each of the state variables of our system\textemdash{}$u$, $w$, $p$, $\Theta$ and $\Phi$\textemdash{}depend upon the independent variables: $x$, $z$, $Z$ and $T_1$. 
	Using this multiple-scales approach, the partial derivatives become
	\begin{equation}
	\dfrac{\partial }{\partial t} \mapsto \epsilon^2 \dfrac{\partial}{\partial T_1} \qquad \text{ and } \qquad \dfrac{\partial }{\partial z} \mapsto \dfrac{\partial}{\partial z} + \epsilon \dfrac{\partial}{\partial Z}.
	\end{equation}
	Introducing the notation $\mathbf{\Psi} = (u,w,p,\Theta,\Phi)^T$, we can express each of the variables as a perturbation expansion in $\epsilon$ about the conduction state {$\mathbf{\Psi}_0 = (0,0,0,0,0)^T$}:
	\begin{equation}
	\label{eq:exppsi}
	\mathbf{\Psi} = \mathbf{\Psi}_0 + \epsilon \mathbf{\Psi}_1 + \epsilon^2 \mathbf{\Psi}_2 + \dots,
	\end{equation}
	where $\mathbf{\Psi}_j = (u_j,w_j,p_j,\theta_j,\phi_j)^T$ is the correction at $O(\epsilon^j)$  for $j = 1,2,\dots$.

	The corrections are periodic in $z$ and also satisfy homogeneous boundary conditions at each order in $\epsilon$:
	\begin{equation}
	u_j = w_j = \theta_j = \phi_j = 0 \qquad \text{ on } x = 0,1, \quad j = 1,2,\dots, \label{eq:lin_bc}
	\end{equation}
	as well as the pressure boundary condition:
	\begin{equation}
	\frac{\partial^2 u_j}{\partial x^2} - \frac{\partial p_j}{\partial x} = 0 \qquad \text{ on } x = 0,1, \quad j = 1,2,\dots.
	\end{equation}
	The expansion (\ref{eq:exppsi}) is substituted into the full system (\ref{eq:NS_conv}\textendash\ref{eq:C_conv}) and the perturbations are solved numerically order-by-order in $\epsilon$ using an extension of the aforementioned collocation method.
	By further extracting the parameter dependence of the perturbations at each order, we obtain a Ginzburg\textendash{}Landau equation that can be applied for all parameter values and will indicate the criticality of the primary bifurcation. 

	We proceed by detailing this formulation, which should be applied to the cases $Le > 1$ and $Le < 1$ separately, owing to the parameter combination $Ra(1-Le)$ changing sign between them.
	However, the results for $Le <1$ can be related to those for $Le>1$ by using an alternative non-dimensionalisation to (\ref{eq:nondim}) involving the solutal diffusivity, $D$, instead of thermal diffusivity, $\kappa$, which results in a set of equations like (\ref{eq:NS}\textendash{}\ref{eq:C}), except with $T$ and $C$ exchanged and the Lewis, Prandtl and Rayleigh numbers replaced by the inverse Lewis number, Schmidt number, $Sc = LePr$, and solutal Rayleigh number, $Ra_S = -RaNLe$, respectively.
	
	The conduction state solves the system at leading order. 
	At $\mathcal{O}(\epsilon)$, the correction is given by the solution to linear system (\ref{eq:linsystem}):
	\begin{equation}
	\mathbf{\Psi}_1 = A_1(Z,T_1) \bigg(U_1(x),W_1(x),P_1(x),\Theta_1(x),Le \Theta_1(x)\bigg)^Te^{ik_cz} + c.c.,\label{eq:O(eps)}
	\end{equation}
	where $k_c$ is the critical wavenumber. 
	No phase constraint is applied at this point, but the amplitude of the eigenfunction is fixed using:
	\begin{equation}
	\langle U_1\,,\,U_1\rangle + \langle W_1\,,\,W_1\rangle+ \langle P_1\,,\,P_1\rangle + \langle \Theta_1\,,\,\Theta_1\rangle = 1,
	\end{equation}
	with the inner product:
	\begin{equation}
	\langle f\,,\,g\rangle = \frac{1}{\lambda_c}\int_0^{\lambda_c}\int_0^1 \overline{f}^T g\,dx\,dz, \label{eq:inn_prod}
	\end{equation}
	where $\lambda_c = 2\pi/k_c$ is the wavelength of the critical eigenvector, the overbar denotes complex conjugation and the superscript $T$ denotes the transposition operation when $f$ is a vector. 
	Due to the lack of available explicit expressions for the solutions to this perturbation problem, these inner products need to be computed numerically, which we achieved by using the Clenshaw\textendash{}Curtis quadrature on the collocation nodes used in Section \ref{sec:linstab}.
	The amplitude $A_1$ evolves over both long spatial and temporal scales according to an amplitude equation that will be determined at higher order.
	
	At $\mathcal{O}(\epsilon^2)$, the linear operator $\mathcal{L}$ acts on the second-order terms and is forced by both the nonlinear terms between the $\mathcal{O}(\epsilon)$ corrections and terms proportional to the slow spatial derivative of the $O(\epsilon)$ correction $A_{1Z}$:
	\begin{align}
	\mathcal{L}(Ra_c)\mathbf{\Psi}_2 &=\underbrace{\begin{pmatrix}
		\displaystyle\frac{1}{Pr}f_{10} |A_1|^2 + \left(A_{1Z}f_{11}e^{ik_cz} + c.c.\right) + \displaystyle\frac{1}{Pr}\left(f_{12} A_1^2e^{2ik_cz} + c.c.\right)\\
		\displaystyle\frac{1}{Pr}f_{20} |A_1|^2 + \left(A_{1Z}f_{21}e^{ik_cz} + c.c.\right)+\displaystyle\frac{1}{Pr}\left(f_{22} A_1^2e^{2ik_cz} + c.c.\right)\\
		\hphantom{f_{20} |A_1|^2 +}\left(A_{1Z}f_{31}e^{ik_cz} + c.c.\right)\hphantom{+\left(f_{22} A_1^2e^{2ik_cz} + c.c.\right)}\\
		f_{40} |A_1|^2 + \left(A_{1Z}f_{41}e^{ik_cz} + c.c.\right)+\left(f_{42} A_1^2e^{2ik_cz} + c.c.\right)\\
		Lef_{40} |A_1|^2	+ \left(A_{1Z}f_{41}e^{ik_cz} + c.c.\right)+ Le\left(f_{42} A_1^2e^{2ik_cz} + c.c.\right)\\
		\end{pmatrix}}_{\displaystyle\mathcal{N}_2}, \label{eq:O(eps^2)}
	\end{align}
	where the functions $f_{ij}(x)$ for $i = 1,2,3,4$ and $j = 0,1,2$ are independent of $Pr$ and $Le$ and are given in table \ref{tab:f_ij}.
	
		\begin{table*}
		\centering
		\begin{tabular}{rrlll}
			\toprule
			\multicolumn{2}{l}{\multirow{2}{*}{$f_{ij}$}}	 & &$j$ & \\ 
			&  & $0$&$1$ &$2$ \\
			\midrule
			&$1$	& $\overline{U}_1\frac{dU_1}{dx} + ik_c\overline{W}_1U_1 + c.c.$ & $-2ik_cU_1$ & $U_1\frac{dU_1}{dx} + ik_cW_1U_1$  \\ 
			&	$2$	& $\overline{U}_1\frac{dW_1}{dx} + ik_c\overline{W}_1W_1 + c.c.$ & $-2ik_cW_1 + P_1$ & $U_1\frac{dW_1}{dx} + ik_cW^2_1$ \\ 
			$i$	&$3$	& $0$ & $-W_1$ & $0$ \\
			& $4$& $\overline{U}_1\frac{d\Theta_1}{dx} + ik_c\overline{W}_1\Theta_1 + c.c.$& $-2ik_c\Theta_1$ & $U_1\frac{d\Theta_1}{dx} + ik_cW_1\Theta_1$\\
			\bottomrule
		\end{tabular} 
	\caption{	Functions $f_{ij}$ ($i = 1,2,3,4$, $j = 0,1,2$) in the nonlinear term $\mathcal{N}_2$ at $\mathcal{O}(\epsilon^2)$ in (\ref{eq:O(eps^2)}). 
			The overbar denotes complex conjugation.}
		\label{tab:f_ij}
	\end{table*}
	
	To ensure the existence of a unique solution at this order, we derive a solvability condition using the Fredholm alternative theorem.
	This involves the adjoint operator to $\mathcal{L}$, $\mathcal{L}^{\dagger}$, defined through the relationship
	\begin{equation}
	\langle f\,,\,\mathcal{L}g\rangle = \langle \mathcal{L}^{\dagger}f\,,\,g\rangle, \label{eq:def_adj}
	\end{equation}
	which holds for all vector functions $f$ and $g$.
	Integrating the left-hand side by parts, we find that the adjoint operator takes the form:
	\begin{equation}
	\mathcal{L}^{\dagger} = \begin{pmatrix}
	\nabla^2 & 0 & -\partial_x & -1 & -1 \\
	0 & \nabla^2 & -\partial_z & 0 & 0 \\
	\partial_x & \partial_z & 0 & 0 & 0 \\
	0 & Ra_c & 0 & \nabla^2 & 0\\
	0 & -Ra_c & 0 & 0 & \frac{1}{Le}\nabla^2 \\
	\end{pmatrix},
	\end{equation}
	together with the adjoint boundary conditions:
	\begin{align}
	u^{\dagger} = 0,\,w^{\dagger} = 0,\,\theta^{\dagger} = 0,\, \phi^{\dagger} = 0 \quad& \text{ on } x = 0,1, \\
	\frac{\partial^2 u^{\dagger}}{\partial x^2} -\frac{\partial p^{\dagger}}{\partial x} = 0\quad& \text{ on } x = 0,1,\label{eq:adj_bc}
	\end{align}
	and periodicity in the vertical direction.
	
	The Fredholm alternative allows us to pose the adjoint problem:
	\begin{equation}
	\mathcal{L}^{\dagger} \mathbf{\Psi}^{\dagger} = \mathbf{0}, \label{eq:adj}
	\end{equation}
	whose solution is unique up to a vertical translation and a multiplicative constant. This solution may be written in the form:
	\begin{equation}
	\mathbf{\Psi}^{\dagger} = \left(U^{\dagger}(x), W^{\dagger}(x),P^{\dagger}(x), \frac{1}{1-Le}\Theta^{\dagger}(x),-\frac{Le}{1-Le}\Theta^{\dagger}(x) \right)^Te^{ik_cz} + c.c., \label{eq:dep_adj}
	\end{equation}
	where the parameter dependence of the components has been extracted. The amplitude and phase are fixed by imposing the conditions:
	\begin{equation}
	\langle U^{\dagger}\,,\,U^{\dagger}\rangle + \langle W^{\dagger}\,,\,W^{\dagger}\rangle+ \langle P^{\dagger}\,,\,P^{\dagger}\rangle + \langle \Theta^{\dagger}\,,\,\Theta^{\dagger}\rangle  = 1,
	\end{equation}
	and
	\begin{equation}
	Im\left(\langle U^{\dagger}\,,\,U_1\rangle\right) = 0,
	\end{equation}
	where $Im$ represents the imaginary part, respectively. 
	
	Using this adjoint solution, we then apply the $\mathcal{O}(\epsilon^2)$ solvability condition:
	\begin{equation}
	\langle \Psi^{\dagger}\,,\,\mathcal{N}_2\rangle = 0. \label{eq:solv_O2_2}
	\end{equation}
	Owing to the vertical wavenumber dependence of terms in $\mathcal{N}_2$, the only non-trivial contributions come from those proportional to $A_{1Z}$ and their complex conjugates 
	and equation (\ref{eq:solv_O2_2}) then reduces to
	 \begin{equation}
	 -2ik_c\langle U^{\dagger},U_1\rangle -2ik_c\langle W^{\dagger},W_1\rangle + \langle W^{\dagger},P_1\rangle - \langle P^{\dagger},W_1\rangle -2ik_c \langle \Theta^{\dagger},\Theta_1\rangle= 0, \label{eq:solv_O2_exp}
	 \end{equation} 
	 which may be further simplified to: 
	\begin{equation}
	\left\langle \mathbf{\Psi}^\dagger\,,\,\dfrac{\partial \mathcal{L}\mathbf{\Psi}}{\partial k_c}\right\rangle = 0.
	\end{equation}
	Thus, this solvability condition is automatically satisfied as the primary bifurcation occurs at a quadratic minimum of the marginal stability curve (see figure~\ref{fig:singlemargstabcurve}). 
	
	The $\mathcal{O}(\epsilon^2)$ system (\ref{eq:O(eps^2)}) can be solved to find that the second-order correction to the conduction state is
	\begin{equation}
	\Psi_2 = |A_1|^2\Psi_2^0 + ((A_2\Psi_1^1 +A_{1Z}\Psi_2^1)e^{ik_cz} + c.c.) + (A_1^2\Psi_2^2e^{2ik_cz} + c.c.),\label{eq:dep_2}
	\end{equation}
	where $\Psi_2 = (u_2,w_2,p_2,\theta_2,\phi_2)^T$ and the functions $\Psi_2^i$ for $i=0,1,2$ have the following parameter dependence:
	\begin{align}
	\mathbf{\Psi}_2^0 &= \left(0,\frac{1}{Pr}\tilde{w}_2 + (1+Le)\tilde{w}_3,\frac{1}{Pr}\tilde{p}_2,\tilde{\theta}_3, Le^2\tilde{\theta}_3\right)^T,\label{eq:dep_psi_0}\\
	\mathbf{\Psi}_2^1 &= \bigg(\tilde{u}_7,\tilde{w}_7,\tilde{p}_7,\tilde{\theta}_7,Le\,\tilde{\theta}_7\bigg)^T,\\
	\begin{split}\mathbf{\Psi}_2^2 &= \left(\frac{1}{Pr}\left(\tilde{u}_4,\tilde{w}_4,\tilde{p}_4,\tilde{\theta}_4,Le\tilde{\theta}_4\right)+(1+Le)(\tilde{u}_5,\tilde{w}_5,\tilde{p}_5,0,0) \right.\\& \qquad \qquad+ (0,0,0,\tilde{\theta}_5,Le^2\tilde{\theta}_5)+Le(0,0,0,\tilde{\theta}_6, \tilde{\theta}_6)\bigg)^T.\end{split}\label{eq:dep_psi_2}
	\end{align}
	The newly introduced functions $\tilde{u}_i$, $\tilde{w}_i$, $\tilde{p}_i$ and $\tilde{\theta}_i$ for $i = 2,...,7$ are independent of $Le$ and $Pr$ and satisfy equations (\ref{eq:O2_0}\textendash{}\ref{eq:bc0O2}) in Section~\ref{sec:O2corrections} of the Appendix.

	Continuing to $\mathcal{O}(\epsilon^3)$, both the deviation away from the critical Rayleigh number and the slow time-dependence of the solution appear in the right-hand side of the resulting system, in addition to nonlinear terms between first and second-order corrections and terms with slow spatial derivatives. 
	The system to solve at third-order is
	\begin{equation}
	\mathcal{L}(Ra_c)\mathbf{\Psi}_3 = \underbrace{\begin{pmatrix}\frac{1}{Pr}\left(\frac{\partial u_1}{\partial_{T_1}} +
		J(\mathbf{u},u) \right) -2\frac{\partial^2u_2}{\partial z\partial Z} - \frac{\partial^2u_1}{\partial Z^2}\\
		\frac{1}{Pr}\left(\frac{\partial w_1}{\partial_{T_1}}  +
		J(\mathbf{u},w)\right) 
		-r\left(\theta_1 - \phi_1\right) -2\frac{\partial^2w_2}{\partial z\partial Z} - \frac{\partial^2w_1}{\partial Z^2}
		+\frac{\partial p_2}{\partial Z}\\
		{}-\frac{\partial w_2}{\partial Z}\\
		\left(\frac{\partial \theta_1}{\partial_{T_1}} 
		+J(\mathbf{u},\theta) \right)
		-2\frac{\partial^2\theta_2}{\partial z\partial Z} - \frac{\partial^2\theta_1}{\partial Z^2} \\
		\left(\frac{\partial \phi_1}{\partial_{T_1}} 
		+J(\mathbf{u},\phi)\right)
		-\frac{2}{Le}\frac{\partial^2\phi_2}{\partial z\partial Z} - \frac{1}{Le}\frac{\partial^2 \phi_1}{\partial Z^2}                                                                                                                                                              
		\end{pmatrix}}_{\mathcal{N}_3},\label{eq:O(eps3)}
	\end{equation}
	where the advective terms are
	\begin{equation}
	J(\mathbf{u},f) = \mathbf{u}_1\cdot \nabla f_2 + \mathbf{u}_2\cdot \nabla f_1 + w_1\partial_Zf_1,
	\end{equation}
	where $f_1$ and $f_2$, respectively, refer to the first- and second-order corrections of the variables $f= u,w,\theta$ and $\phi$.
	
	The solvability condition at this order:
	\begin{equation}
	\langle \mathbf{\Psi}^{\dagger} \;,\;\mathcal{N}_3\rangle = 0, \label{eq:O3solvability}
	\end{equation}
	is no longer trivially satisfied because some nonlinear terms contained in $\mathcal{N}_3$ have $e^{ik_cz}$ dependence arising from terms proportional to $A_1$, $|A_1|^2A_1$, $A_{1ZZ}$, $A_{2Z}$ and their complex conjugates.
	However, the contributions to (\ref{eq:O3solvability}) from terms proportional to $A_{2Z}$, cancel for the same reason that the solvability condition at $O(\epsilon^2)$ was satisfied.
	Consequently, $A_2$ remains arbitrary at this order. 
	Collecting the remaining terms from equation (\ref{eq:O3solvability}), we obtain the Ginzburg\textendash{}Landau equation (GLE), that holds for both $Le > 1$ and $Le < 1$:
	\begin{equation}
	\alpha A_{1T_1} = \gamma r A_1 + \beta |A_1|^2A_1 + \delta A_{1ZZ}, \label{eq:GLE1}
	\end{equation}
	where table~\ref{tab:solv_O3} indicates which terms of $\mathcal{N}_3$ contribute to each term above.
	This equation is equivariant under the $O(2)$ symmetry so we may chose the phase of the $\mathcal{O}(\epsilon)$ correction so that these coefficients are real.
	The coefficient $\delta$ is independent of the physical parameters $Pr$ and $Le$, while $\alpha, \beta$ and $\gamma$ satisfy the relations:
	\begin{align}
	\alpha &= \frac{1}{Pr}\alpha_1 + (1+Le)\alpha_2,\label{eq:alp}\\
	\beta &= \frac{1}{Pr^2}\beta_1 + \frac{1+Le}{Pr}\beta_2 + (1+Le^2)\beta_3+Le\beta_4, \label{eq:bet}\\
	\gamma &= (1-Le)\gamma_1,\label{eq:gam}
	\end{align}
	where full expressions used to obtain $\alpha_i$, $\beta_i$, $\gamma_1$ and $\delta$ are provided in (\ref{eq:alp2}\textendash{}\ref{eq:del}) and evaluated in table \ref{tab:coeff_values} in Section~\ref{sec:GLEcoeff}.
	By dividing (\ref{eq:GLE1}) through by $\alpha$, equation (\ref{eq:GLE1}) is more conveniently written as
	\begin{equation}
	A_{1T_1}= a_1rA_1 + a_2 |A_1|^2A_1 + a_3A_{1ZZ} \label{eq:GLE2},
	\end{equation}
	where $a_1 = \gamma/\alpha$, $a_2 = \beta /\alpha$ and $a_3 = \delta/ \alpha$. 
			\begin{table*}
		\centering
		\begin{tabular}{rcccc}
			\toprule
		&	\multicolumn{4}{c}{Term in the Ginzburg--Landau equation (\ref{eq:GLE1})}	\\
		& \hspace{0.5cm}	$\alpha A_{1T_1}$\hspace{0.5cm} & \hspace{0.5cm}$\gamma r A_1$ \hspace{0.5cm}& \hspace{0.5cm}$\beta |A_1|^2 A_1$ \hspace{0.5cm}&\hspace{0.5cm}$\delta A_{1ZZ}$\hspace{0.5cm} \\
		\midrule
		\multirow{4}{*}{Terms in $\mathcal{N}_3$ proportional to} & \multirow{4}{*}{$\frac{\partial f_1}{\partial T_1}$}&\multirow{4}{*}{ $r(\theta_1-\phi_1)$ }& \multirow{2}{*}{$\mathbf{u}_1\cdot \nabla f_2$}& $\frac{\partial^2 f_2}{\partial z\partial Z}$\vspace{0.1cm}\\
		& & & &  $\frac{\partial f_1}{\partial Z^2}$\\
                & & & \multirow{2}{*}{$\mathbf{u}_2 \cdot \nabla f_1$} & $\frac{\partial p_2}{\partial Z}$ \\
                & & & & $\frac{\partial w_2}{\partial Z}$ \\
			\bottomrule
		\end{tabular} 
		\caption{Terms from $\mathcal{N}_3$ (see equation (\ref{eq:O(eps3)})) contributing to the Ginzburg--Landau equation (\ref{eq:GLE1}). The column in which these terms are placed informs on the term to which they contribute. Here, $f_1$ and $f_2$, respectively refer to first- and second-order corrections of the variables $f= u,w,\theta$ and $\phi$.}
		\label{tab:solv_O3}
	\end{table*}
                        
	The solutions to the Ginzburg\textendash{}Landau equation (\ref{eq:GLE2}) are good approximations of the small-amplitude solutions of the full doubly diffusive system (\ref{eq:NS_conv}\textendash{}\ref{eq:C_conv}).	
	Of particular interest here are the two steady solutions that are invariant with respect to the long spatial scale $Z$.
	The first of these solutions is the trivial solution:
	\begin{equation}
	A_1 = 0. \label{eq:A10}
	\end{equation}
	This solution is valid for all $r$ and corresponds to the conduction state (\ref{eq:conduction_state_2}).
	The second important solution is:
	\begin{equation}
	A_1 = \left(-\frac{a_1 r}{a_2}\right)^{1/2}e^{i\chi}, \label{eq:A1conv}
	\end{equation}
	where $\chi$ is an arbitrary phase.
        This solution relates to states of small-amplitude spatially-periodic convection that can be found near the primary bifurcation. These fluid states can then be approximated by
	\begin{equation}
	\begin{split} \left(u,w,p,\Theta,\Phi\right)^T \approx   \sqrt{-\frac{a_1 (Ra-Ra_c)}{a_2}} \bigg(U_1(x),W_1(x),P_1(x),\Theta_1(x),Le \Theta_1(x)\bigg)^Te^{ik_cz} + c.c,
	\end{split}\label{eq:per_conv}
	\end{equation}
	where the phase $\chi$ has been absorbed into $z$ via a vertical translation.
	These states only exist at small-amplitude for Rayleigh numbers that satisfy
	\begin{equation}
	\frac{a_1}{a_2}\left(Ra_c-Ra\right) > 0.
	\end{equation}
	Consequently, the sign of the ratio $a_1/a_2$ determines the criticality of the primary bifurcation and the initial direction of branching.

\begin{figure}
		\centering
		\includegraphics{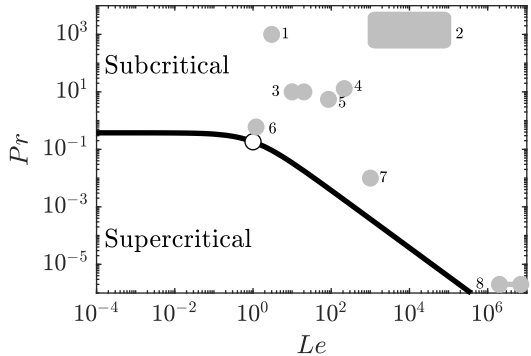}
		\caption{Boundary $a_2 = 0$ in $(Le,Pr)$ parameter space separating the region where the primary bifurcation from the conduction state is subcritical (above) from that where it is supercritical (below). 
		The conduction state is linearly stable for all $Ra$ at $Le = 1$ and this point is indicated by the open circle.
		The grey regions indicate parameter values from \cite{schmitt1983characteristics} for the physical doubly diffusive systems: (1) salt/sugar, (2) magmas, (3) oxide semiconductors, (4) heat/salt $0^{\circ}$C, (5) heat/salt $30^{\circ}$C, (6) humidity/heat, (7) liquid metals and (8) stellar interiors.}
		\label{fig:vertboundary}
		\end{figure}

	Using the numerical values in table~\ref{tab:coeff_values}, we computed $a_i$ over a range of parameter values.
	The coefficients $a_1$ and $a_3$ are positive for all $Pr$ provided $Le \neq1$, whereas the sign of $a_2$ changes as these parameters are varied.
	As a result, there exists a boundary in parameter space that separates regions where the primary bifurcation is subcritical ($a_2 >0$) from those where it is supercritical ($a_2 < 0$).
	This boundary is shown in figure~\ref{fig:vertboundary} and implies that, for any value of the Lewis number, there exists a critical value of the Prandtl number, $Pr_c(Le)$, expressed in terms of the physical parameters in (\ref{eq:Prc}), above which the bifurcation is subcritical.
	This critical value tends to $0.376$ for small Lewis numbers while it approaches the asymptotic relation $Pr_c \sim 0.376/Le$ as the Lewis number tends to infinity.
	We further note that the parameter values for physical doubly diffusive systems from \cite{schmitt1983characteristics} all lie within the region where the primary bifurcation is subcritical.
	While we are unaware of further fluid systems lying within the supercritical region of parameter space, we expect that they exist since some of the physical systems identified in figure~\ref{fig:vertboundary}, including humidity/ heat and stellar interiors (marked (6) and (8), respectively), have parameter values that are within an order of magnitude of the sub/supercritical boundary. 
	
	We can gain physical insight into the criticality of the primary bifurcation by examining the contributions that each of the nonlinear terms from equations (\ref{eq:NS}--\ref{eq:C}) make to $a_2$, using a similar approach to the one \citet{requile2020weakly} applied to plane Poiseuille and plane Couette flows with viscous dissipation.
	The expression of the coefficient $\beta$ (\ref{eq:bet}) and the corresponding numerical values provided in table~\ref{tab:coeff_values} in the Appendix, show that the inertial term $\mathbf{u}\cdot \nabla \mathbf{u}$ (contributing to $\beta_1$ and $\beta_2$) provides a negative contribution to $a_2$, whereas thermal $\mathbf{u}\cdot \nabla T$ and solutal $\mathbf{u}\cdot \nabla C$ advective terms (mostly contributing to $\beta_3$ and $\beta_4$) provide a positive contribution to $a_2$.
	The latter statement is further justified in Section \ref{sec:adv} in the Appendix.
	It is therefore solutal and thermal advection in the system that drives the subcriticality of the primary bifurcation, while inertial effects drive the supercriticality. 
	Thus, reducing the Prandtl number reduces the subcriticality of the bifurcation since the effects of inertia are strengthened.
	
	The final term in the Ginzburg\textendash{}Landau equation (\ref{eq:GLE2}), $a_3A_{1ZZ}$, allows small-amplitude solutions of the doubly diffusive system to exhibit long scale amplitude modulation.
	These solutions include phase-winding states that describe patterns whose wavenumbers are close to the critical wavenumber $k_c$ \citep{Cross1983}, and spatially modulated states that can develop into localised states away from the primary bifurcation \citep{bergeon2008periodic}. These are out of the scope of the present work, but we will consider the effect of the term $a_3A_{1ZZ}$ on spatially localised states in future work.
		
	\section{Fully nonlinear behaviour}\label{sec:nonlinearanalysis}
	
	Having established the region of $(Le,Pr)$ parameter space in which the bifurcation is subcritical, we can now investigate the nonlinear behaviour of the system near the onset of convection.
	In particular, we focus on the structure and stability of the primary branch of spatially periodic convection states with wavenumber $k_c$ as it extends towards larger amplitudes.
	For this, we consider a single-wavelength domain with ${L_z = \lambda_c = 2\pi/k_c}$, which precludes modulational instabilities arising in large domains that are captured by our weakly nonlinear analysis through the $A_{1ZZ}$ term in equation (\ref{eq:GLE2}).

	We numerically continue the primary branch against the Rayleigh number across a range of Lewis ($Le \in [5,100]$) and Prandtl ($Pr \in [2\times10^{-3}, 10]$) numbers.
	Cases for which $Le < 1$ can be extrapolated from our results by a suitable transformation.
	The solution branches will be identified on bifurcation diagrams showing either the total kinetic energy of steady states:
	\begin{equation}
	E = \frac{1}{2}\int_0^{\lambda_c}\int_0^1 \left(\,u^2+w^2\right)\,dx\,dz, \label{eq:ke}
	\end{equation}
	or the average velocity $\|\mathbf{u}\|_2 = \sqrt{2E/\lambda_c}$, against the Rayleigh number $Ra$.
	
	Computations were carried out using a spectral element numerical method based on a Gauss--Lobatto--Legendre discretisation \citep{bergeon2002natural} and supplemented by Stokes preconditioning with $\Delta t = 0.1$, as detailed by \citet{beaume2017adaptive}.
	Numerical results were validated against a discretisation of up to $4$ spectral elements with $29$ nodes in both the $x$ and $z$ directions.
	The stability of the steady states was computed using an Arnoldi method based on a time-stepping scheme \citep{mamun1995asymmetry}.
	Further direct numerical simulations used a stiffly stable second-order splitting scheme based on \cite{karniadakis1991high} with time-step $\Delta t = 10^{-3}$.
	
	\subsection{Bifurcation structure}
	
\begin{figure}
		\centering
		\includegraphics[width=\linewidth]{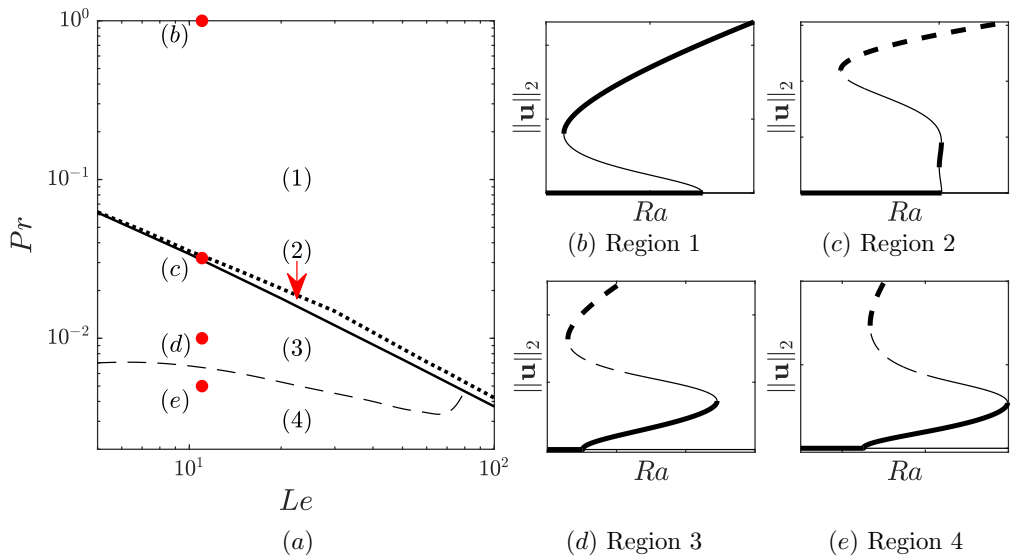}
		\caption{(a) Enlargement of a subset of the parameter space shown in figure \ref{fig:vertboundary} showing four regions where the bifurcation diagrams exhibit qualitatively different behaviours.
		The thick line separates subcritical from supercritical branching, while the additional region boundaries are identified with either dotted or dashed lines. 
		(b--e) Representative bifurcation diagrams for parameter values within each of the four regions.
		The stability of the branch segments are also indicated using thick lines for stable solutions, thin lines for solutions unstable to amplitude perturbations and dashed lines for solutions unstable to drift.
		The location of bifurcations depend upon the specific parameter values used, so those used for each sketch have been marked in panel (a). }
		\label{fig:parregimesbd}
	\end{figure}

	The results can be summarised by dividing parameter space according to the qualitative nature of the bifurcation diagram.
	Figure~\ref{fig:parregimesbd}(a) indicates the four main regimes found.
	Region~(1) describes the moderate and large $Pr$ behaviour for all $Le$.
	In this region, the primary bifurcation is strongly subcritical and the primary branch has a single saddle-node, as shown in figure~\ref{fig:parregimesbd}(b).
	Parameter values within this region have received the most attention in previous studies focusing on subcritical pattern formation (e.g. see \citep{xin1998bifurcation,bergeon2008spatially}).
	Region~(2) occupies a small region of parameter space above the boundary $Pr = Pr_c$, where the primary bifurcation is weakly subcritical, and separates the typical subcritical behaviour in region~(1) from the supercritical behaviour in regions~(3) or (4). 
	The steady convection state branches typically have three saddle-nodes in region (2), as exemplified in figure~\ref{fig:parregimesbd}(c).
	Regions (3) and (4) identify the two qualitatively different types of bifurcation diagrams observable when the primary bifurcation is supercritical.
	In both cases, the primary branch has two saddle-nodes, with the first lying in the supercritical region $Ra > Ra_c$.
	The difference between the regions is the location of the second saddle-node: in region (3), it is found for $Ra < Ra_c$ (see figure~\ref{fig:parregimesbd}(d)), whereas, in region (4), it is found in $Ra > Ra_c$ (see figure~\ref{fig:parregimesbd}(e)).
	Consequently, a large-amplitude convection state may coexist with the stable conduction state when the primary bifurcation is supercritical, but, for sufficiently small $Pr$, steady convection states are found entirely within the supercritical region, where the conduction state is unstable. 
	There may exist a fifth region, where the primary branch increases monotonically in both Rayleigh number and in amplitude, but we have not identified it in this study.
	
\begin{figure}
		\centering
		\includegraphics[]{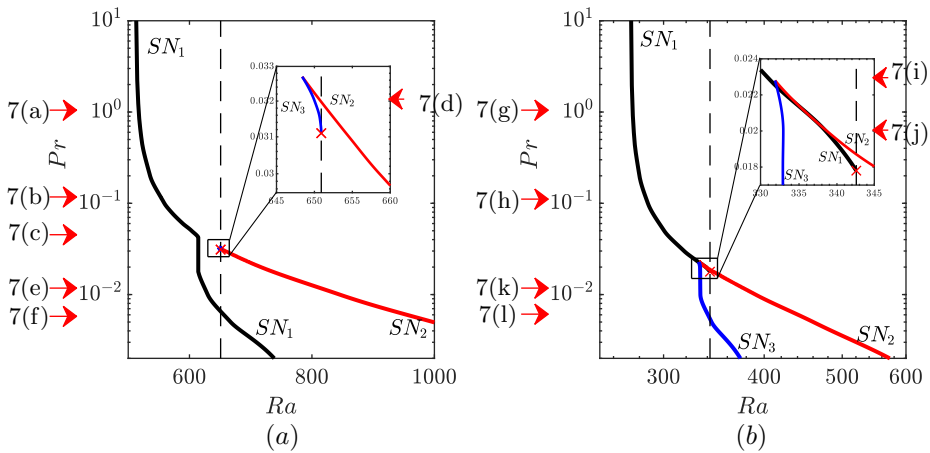}
		\caption{Location of the three saddle-node bifurcations of the primary branch in $(Ra,Pr)$ parameter space for (a) $Le = 11$ and (b) $Le = 20$. The dashed line marks the critical Rayleigh number at which the primary bifurcation is found, $Ra_c$, and the cross marks the codimension two point $(Ra_c,Pr_c)$ explained in the text. The insets provide enlargements of the area around $Ra_c$ in each case. Arrows mark the bifurcation diagrams shown in figure~\ref{fig:bifurcationdiagramstransition}.}
		\label{fig:saddle_nodes}
	\end{figure}

	We now determine the structure of the primary branch as $Pr$ decreases for a fixed value of $Le$.
	To achieve this, we follow the locations of its three saddle-nodes with respect to $Ra$ and $Pr$.
	In doing so, we observed two different scenarios according to whether the pair of saddle-nodes are created on the lower or upper part of the primary branch.
	These are exemplified in figure~\ref{fig:saddle_nodes} for $Le = 11$ (representative of $5 \leqslant Le \lesssim 15$) and $Le = 20$ (representative of $19 \lesssim Le < 100$).
	Since the transition between the two scenarios occupies a small region of parameter space within region (2) for $15 \lesssim Le \lesssim 19$, we did not investigate it any further. 
	
\begin{figure}
		\centering
		\includegraphics[width=\linewidth]{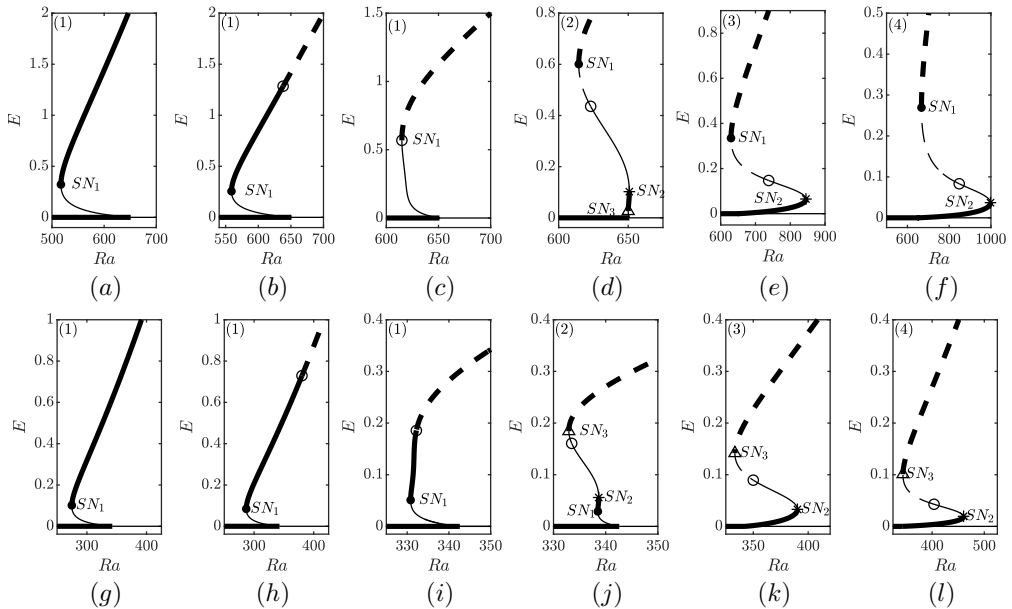}
		\caption{	Bifurcation diagram showing the primary branch of steady convection and the stability of the related states across the four regions, indicated in the top left corner.
			Thick lines indicate stable solutions, thin lines indicate solutions unstable to amplitude perturbations and dashed lines indicate solutions unstable to drift.
		Saddle-nodes are marked by symbols: $SN_1$ (filled circle), $SN_2$ (asterisk) and $SN_3$ (triangle). 		
		The open circle corresponds to the destabilising drift bifurcation. 	
		The parameter values, also indicated by the arrows in figure~\ref{fig:saddle_nodes}, are: $Le = 11$, and (a) $Pr = 1$, (b) $Pr = 0.1$, (c) $Pr = 0.042$, (d) $Pr = 0.032$, (e) $Pr = 0.01$, (f) $Pr = 0.005$; as well as $Le = 20$, and (g) $Pr = 1$, (h) $Pr = 0.1$, (i) $Pr = 0.023$, (j) $Pr = 0.02$, (k) $Pr = 0.01$ and (l) $Pr = 0.005$. 
		For $Le = 11$ (resp. $Le = 20$), $Pr_c \approx 0.031$, $Pr_{cusp} \approx 0.033 $ (resp. $Pr_c \approx 0.018$, $Pr_{cusp}\approx 0.023 $). }
		\label{fig:bifurcationdiagramstransition}
	\end{figure}

	To help interpret the plots in figure~\ref{fig:saddle_nodes}, figure~\ref{fig:bifurcationdiagramstransition} demonstrates the evolution of the bifurcation diagrams as $Pr$ decreases for $Le = 11$ (panels (a--f)) and $Le = 20$ (panels (g--l)).
	The structure of the bifurcation diagrams in region~(1), for high $Pr$, remain similar, as shown in figures~ \ref{fig:bifurcationdiagramstransition}(a) and \ref{fig:bifurcationdiagramstransition}(g).
	From the primary bifurcation, the primary branch extends towards lower Rayleigh numbers and proceeds to turn around at a saddle-node, hereafter referred to as $SN_1$, before heading towards large amplitude convection states at large $Ra$.
	Figure~\ref{fig:saddle_nodes} suggests that, as $Pr\rightarrow \infty$, the location of $SN_1$ tends to a constant Rayleigh number, dependent upon $Le$.
	This figure also shows that $SN_1$ occurs at larger $Ra$ as the Prandtl number is decreased and the primary bifurcation becomes less subcritical.
	
	Upon decreasing the Prandtl number, the primary branch undergoes a cusp bifurcation at $Pr \approx Pr_{cusp}(Le)> Pr_c(Le)$, while still subcritical, denoting the beginning of region~(2).
	The cusp produces two additional saddle-nodes along the primary branch: $SN_2$ and $SN_3$.
	The exact process by which this is achieved depends on the Lewis number.
	For $Le \lesssim 15$, the cusp bifurcation occurs at smaller amplitude than $SN_1$ and the saddle-nodes are labelled $SN_3$, $SN_2$, $SN_1$ as the branch is followed in the direction of increasing energy (see, for example, figure~\ref{fig:bifurcationdiagramstransition}(d) for $Le = 11$ and $Pr = 0.032$, near the cusp parameter value: $Pr_{cusp}\approx 0.033$).
	In contrast, for $Le \gtrsim 19$, the cusp bifurcation occurs at higher amplitude than $SN_1$ and saddle-nodes are labelled $SN_1$, $SN_2$, $SN_3$, as shown in figure~\ref{fig:bifurcationdiagramstransition}(i) for $Le = 20$, $Pr = 0.023 \approx Pr_{cusp}$. 
	
	Continuing to reduce $Pr$ across region (2) (from $Pr_{cusp}$ to $Pr_c$), the Rayleigh number associated with $SN_2$ increases so that it reaches the supercritical region before $Pr = Pr_c$.
	During this transition, the saddle-node with smallest amplitude ($SN_3$ for $Le \lesssim 15$; $SN_1$ for $Le \gtrsim 19$) moves to larger Rayleigh numbers but with decreasing amplitude until it collides with the primary bifurcation at ${Pr = Pr_c}$ and ${Ra = Ra_c}$, where the primary bifurcation changes from subcritical to supercritical.
	This process is highlighted in the insets of figure~\ref{fig:saddle_nodes} and results in the primary branch possessing only two saddle-nodes in the supercritical regime ($Pr < Pr_c$).
	
	The locations of the remaining two saddle-nodes go toward larger $Ra$ as $Pr$ decreases and are found in the supercritical region ($Ra > Ra_c$) in region (4), as shown in figure~\ref{fig:parregimesbd}.
	It is therefore clear that multiple steady convection states can exist for the same parameter values near the onset of convection, regardless of the criticality of the primary bifurcation.
	This result extends earlier observations on the number of saddle-node bifurcations occurring along the primary branch in related systems \citep{tsitverblit1993multiplicity}.
	
\begin{figure}	\includegraphics{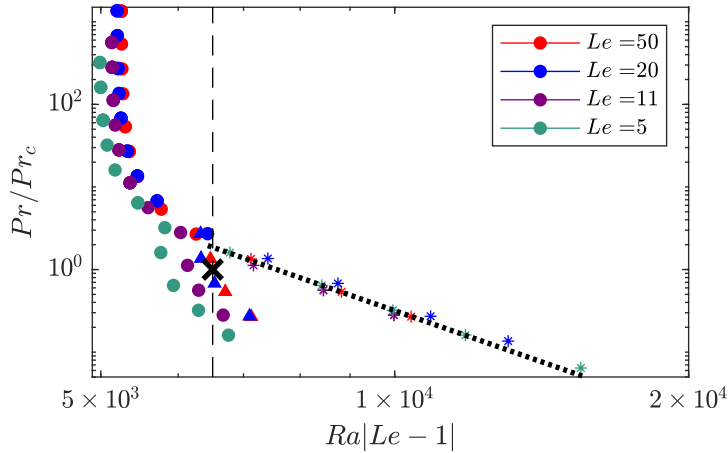}
		\caption{Saddle-node locations for $Le = 5$ (green), $Le=11$ (purple), $Le=20$ (blue) and $Le=50$ (red). The black dashed line marks the location of the primary bifurcation and the red cross marks the codimension two point where the criticality of the primary bifurcation changes, at $Pr=Pr_c$. The black dashed line represents relationship (\ref{eq:RaSN2relationship}). Saddle-nodes are marked by circles for $SN_1$ asterisks for $SN_2$ and triangles for $SN_3$.}
		\label{fig:saddle_nodes_all}
	\end{figure}

	More insight into these results can be obtained by representing, as in figure~\ref{fig:saddle_nodes_all}, the location of the saddle-nodes for various Lewis numbers as a function of the reduced Prandtl number $Pr/Pr_c$ and combined parameter $Ra|Le-1|$.
	These reduced parameters allows us to identify the location where the criticality of the primary bifurcation changes as the single coordinate point: $Pr/Pr_c = 1$, $Ra|Le-1|\approx  6509$.
	
	Figure \ref{fig:saddle_nodes_all} shows that, for $Pr < Pr_c$ and the chosen values of the Lewis number, the location of the first supercritical saddle-node $SN_2$ can be approximated by:
	\begin{equation}
	Ra_{SN_2} \approx \frac{6460}{|Le-1|}\left(\frac{Pr}{Pr_c}\right)^{-0.24}. \label{eq:RaSN2relationship}
	\end{equation}
	For $Pr < 10^{-2}$ (not shown), the location of saddle-node $SN_2$ deviates from the relation above, indicating a potentially different asymptotic regime.
	These results also illustrate the large $Pr$ behaviour of the subcritical saddle-node $SN_1$: $Ra_{SN_1}|Le - 1|$ tends to a constant as the Prandtl number tends to infinity.
	This constant increases with $Le$ and saturates for large values of the Lewis number.
	These results echo those obtained in doubly diffusive convection in a 2D vertical porous enclosure, where \citet{mamou1998double} used a parallel flow approximation to demonstrate that the Rayleigh number at which the subcritical saddle-node occurs is proportional to $1/(1-Le)$ for large enough Lewis numbers. 
	
	\subsection{Solution profiles}
	
\begin{figure}
		\centering
		\includegraphics[width=\linewidth]{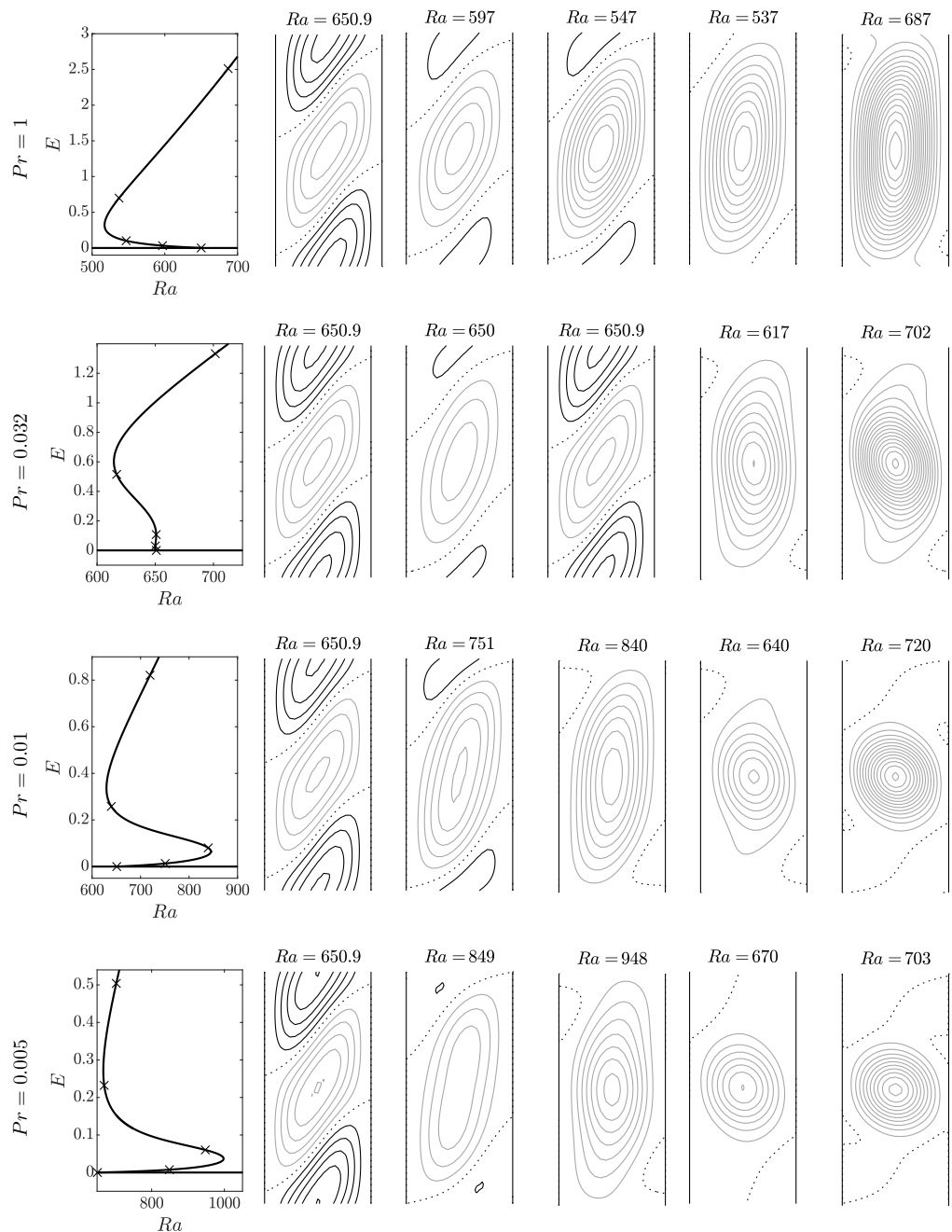}
		\caption{Streamfunctions of the steady states on the primary branch when ${Le = 11}$ for different values of the Prandtl number: top row ${Pr = 1}$, second row ${Pr = 0.032}$, third row ${Pr = 0.01}$ and bottom row ${Pr = 0.005}$. The left column shows the respective bifurcation diagrams and indicates with a cross the solutions that have been represented in the subsequent panels. 
		Black (grey, dotted) contours indicate positive (negative, zero) values of the streamfunction.
		Contour intervals: first column, top two rows\textemdash{}$10^{-4}$; first column, third row\textemdash{}$10^{-5}$; first column, bottom row\textemdash{}$2\times10^{-5}$; second column, top two rows\textemdash{}$0.02$; second column, bottom two rows\textemdash{}$0.01$; third column\textemdash{}$0.02$; fourth and fifth columns\textemdash{}$0.05$.}
		\label{fig:Le11prof}
	\end{figure}

	Despite the different scenarios obtained at different values of the Prandtl number (see figure~\ref{fig:parregimesbd}), the steady convection states undergo similar structural changes along their branch, as evidenced in figure~\ref{fig:Le11prof} for $Le = 11$ and $Pr = 1,\,0.032,\,0.01$ and $0.005$.
	The streamfunction profiles are similar near the primary bifurcation regardless of the value of the Prandtl number (see second column of figure~\ref{fig:Le11prof}), which is in agreement with the linear stability results from figure~\ref{fig:O(eps)soln}(c).
	Moving along the branches in the direction of increasing energy, the first change that we observe is the strengthening of the anticlockwise roll, where fluid near the hotter wall moves upwards.
	This occurs in both the subcritical and the supercritical regimes, as can be seen in the third column of figure~\ref{fig:Le11prof}.
	Continuing the branches to the large-amplitude saddle-node and beyond, the amplitude of the weaker roll decreases, leaving room for the stronger roll to straighten.
	At large enough amplitude, an anticlockwise roll occupies the domain, irrespective of the value of $Pr$.
	Its amplitude grows as the upper branch is followed to larger values of $Ra$, where the Prandtl number starts to impact the flow: the roll occupies a smaller area at lower values of the Prandtl number, as seen within the final column of figure~\ref{fig:Le11prof}.
	This resembles the fly-wheel convection, with nearly circular streamlines, seen in low-Prandtl Rayleigh\textendash{}B\'enard convection as studied by \cite{clever1981low}.
	
\begin{figure}
		\centering
		\includegraphics[width=\linewidth]{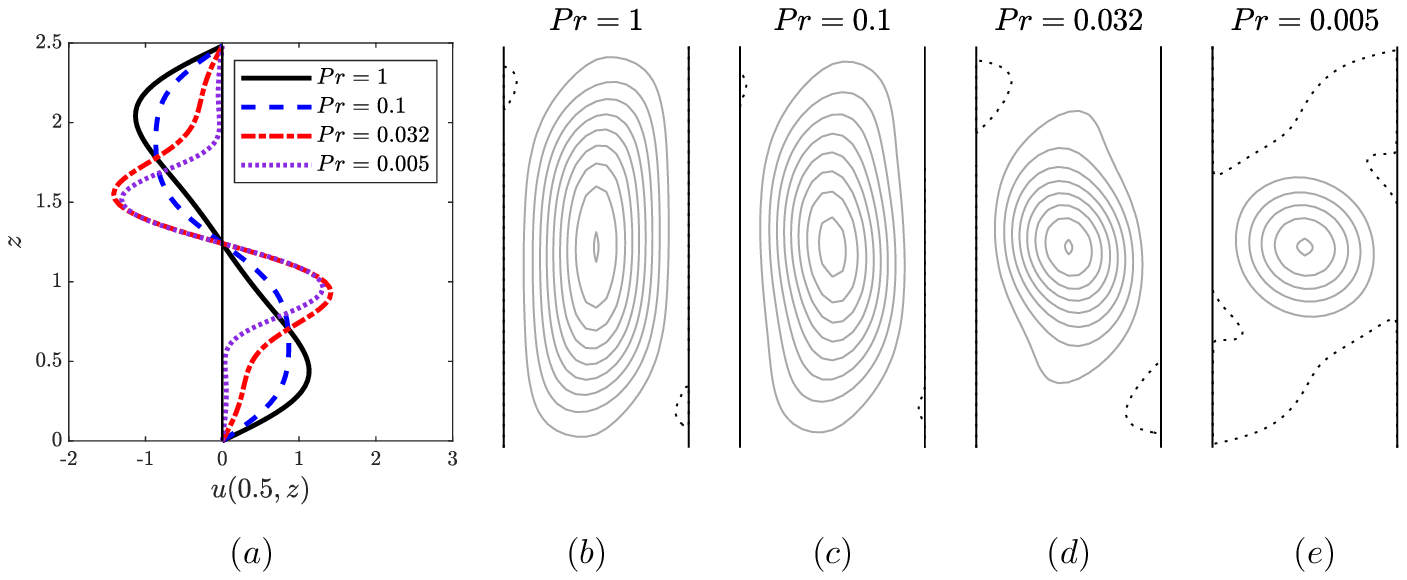}
		\caption{Horizontal velocity and streamfunction of solutions from the upper segment of the primary branch at $Ra = 700$ for $Pr = 1,\,0.1,\,0.032,\,0.005$ and $Le = 11$ represented via (a) the midline horizontal velocity ($u(x=0.5,z)$) and streamfunction contours plots for (b) $Pr = 1$, (c) $Pr = 0.1$, (d) $Pr = 0.032$ and (e) $Pr = 0.005$ with contour intervals $0.1$.}
		\label{fig:Le11horvel700}
	\end{figure}

	To characterise these observations in more detail, figure~\ref{fig:Le11horvel700} reports the horizontal velocity profiles observed on the upper branch for $Pr = 1$, $0.1$, $0.032$ and $0.005$.
	The decrease in roll size is apparent when $Pr$ is decreased.
	This is particularly evident for $Pr = 0.005$, where the horizontal velocity remains small except within the range $0.6 \lesssim z \lesssim 1.9$, in such a way that the roll only occupies about half of the domain's extent.
	Figure~\ref{fig:Le11horvel700}(a) additionally shows the transition to these states from the large rolls observed at $O(1)$ Prandtl numbers.
	For $Pr = 1$, the maximum horizontal velocity is achieved far from the center of the roll, at $z \approx  0.44,\,2.04$, producing a region of strong shear between the rolls and gentle quasi-linear velocity variations inside the rolls.
	As $Pr$ is lowered, these maxima move towards the centre of the roll by initially becoming less pronounced and creating flatter extrema (see figure \ref{fig:Le11horvel700}(c)), followed by the emergence of peaks at $z = 1$ and $z\approx1.5$.
	The maximum horizontal velocity does not change significantly within this range of Prandtl number values in such a way that the low $Pr$ rolls represent narrow regions of strong shear surrounded by low amplitude flow.

	\subsection{Stability of the nonlinear states}\label{sec:stab_tempdyn}
	
	The stability of states on the primary branch is controlled by two eigenmodes: an amplitude mode that preserves the $S_{\Delta}$ symmetry of the system and a drift mode that breaks the $S_{\Delta}$ symmetry.
	The translation mode, associated with vertical translations due to the symmetry $T_{\delta}$, remains marginal along the branch and none of the other eigenmodes become destabilising over the range of parameters considered.
	
	Close to the onset of convection, the amplitude mode is initially destabilising when the bifurcation is subcritical ($Pr > Pr_c$), whereas it is stabilising when the bifurcation is supercritical ($Pr < Pr_c$).
	This mode subsequently changes stability at successive saddle-nodes.
	In particular, it becomes stabilising at saddle-nodes $SN_1$ and $SN_3$, where the branch turns towards higher $Ra$, but becomes destabilising at $SN_2$, where the branch turns towards lower $Ra$.
	As a result, the upper branches of steady convection states are always stable to amplitude perturbations for all $Le$ and $Pr$.
	
\begin{figure}
		\centering
		\includegraphics[]{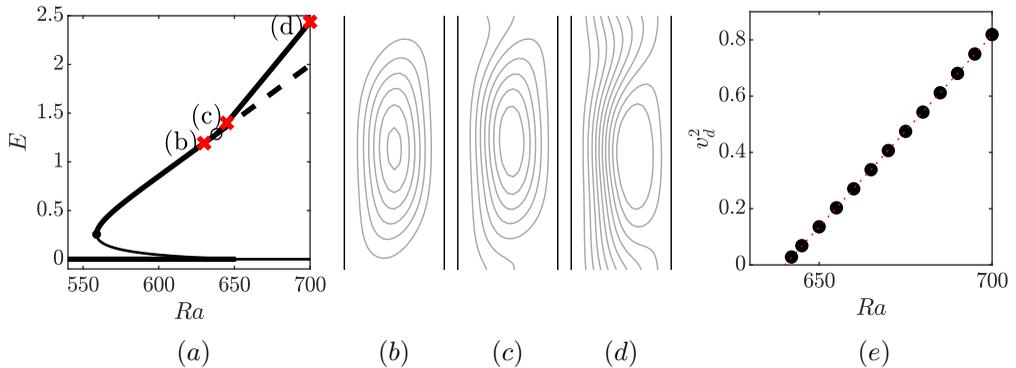}
		\caption{Drift bifurcation and downward travelling waves for $Pr = 0.1$, $Le =11$, for which $Ra_d \approx 638$. (a) Bifurcation diagram showing the kinetic energy $E$ as a function of the Rayleigh number $Ra$ for steady states and travelling waves. Thick lines indicate stable solutions, thin lines indicate solutions unstable to amplitude perturbations and dashed lines indicate solutions unstable to drift. The drift bifurcation is shown by the open circle. (b) Stable convection state at $Ra = 630$ shown by contours of its streamfunction with intervals $0.1$ (first red cross on panel (a)). Further panels show similar representations of stable travelling waves at: (c) $Ra = 645$ and (d) $Ra = 700$. (e) Squared drift speed along the stable branch as a function of the Rayleigh number. The dotted line shows the fitting law: $v_d \approx 0.12\sqrt{Ra - 640}$. }
		\label{fig:travstate}
	\end{figure}

	The drift mode is stabilising near the primary bifurcation at $Ra = Ra_c$ for all $Pr$, but becomes destabilising at a drift-pitchfork bifurcation further along the branch at $Ra = Ra_d$, whose location depends upon both $Le$ and $Pr$, as can be seen in figure~\ref{fig:bifurcationdiagramstransition}.
	The marginal mode is identical to the translation mode at this bifurcation and its destabilisation leads to a pair of branches of travelling wave solutions, as shown in figure \ref{fig:travstate}(a) for $Pr = 0.1$ and $Le = 11$.
	Close to their onset, these states take the form of a single large-amplitude convection roll (see figure~\ref{fig:travstate}(c)) that slowly drifts either upwards or downwards. 
	As these  branches are followed beyond the drift bifurcation, an asymmetric streaming flow strengthens while the convection roll weakens and moves toward the wall where the streaming flow is the weakest.
	This transition is shown from figure~\ref{fig:travstate}(b) at $Ra = 630$ to figure \ref{fig:travstate}(d) at $Ra= 700$.
	At the same time, the drift speed increases at a rate approximately proportional to $\sqrt{Ra-Ra_d}$, as shown in figure \ref{fig:travstate}(e).
	This result extends the findings obtained for $Le = 1.2$, $Pr = 1$ by \citet{xin1998bifurcation} to a wider range of parameter values.
	
	The stability of the travelling waves is determined by the location of the drift bifurcation: these states are initially stable when the bifurcation occurs on the upper branch of steady convection states, whereas they are unstable when the bifurcation occurs along the lower branch.
	Both cases can be achieved for a given $Le$ when $Pr$ is varied, as figure~\ref{fig:bifurcationdiagramstransition} illustrates for selected values of the Prandtl number with $Le = 11$ and $Le = 20$.
	For large values of the Prandtl number, the drift bifurcation occurs on the upper branch at large Rayleigh numbers. 
	This location moves closer to the saddle-node with decreasing Prandtl numbers so that the two coincide at $Pr = Pr^*$ and $Ra = Ra^*$.
	For $Le = 11$, we found that $Pr^*\approx 0.042$ and $Ra^* \approx 614.9$ (see figure~\ref{fig:bifurcationdiagramstransition}(c) for a bifurcation diagram at similar values of the parameters).
	For smaller values of the Prandtl number, the drift bifurcation occurs along the lower branch of convection states and at a value of the Rayleigh number that increases as $Pr$ is decreased.
	For all the parameter values tested, this bifurcation was found to occur at larger amplitude than saddle-node $SN_2$ and, consequently, the small-amplitude steady convection states remain stable to drift.

	\subsection{Dynamical attractors}
	
	The temporal dynamics of the system change as the drift bifurcation passes below the subcritical saddle-node since all the steady convection states from the upper branch and the travelling wave states are destabilised in the process. 
	Many initial conditions will consequently decay towards the conduction state at low $Pr$ and $Ra$. 
	This decay is not possible when the conduction state is unstable for $Ra > Ra_c$, where we find that the dynamics converge on time-dependent states. 
	
	To understand how this behaviour arises, we unfold the saddle-node-pitchfork normal form near the codimension two point $(Ra^*,Pr^*)$ where the drift bifurcation and saddle-node coincide. 
	This unfolding takes the form \citep{guckenheimer1983nonlinear}:
	\begin{align}
	\dot{x} &= -\mu_1 x + b_1 xz,\label{eq:unfoldingx}\\
	\dot{z} &= \mu_2 - x^2 - z^2 + b_2 z^3, \label{eq:unfoldingz}
	\end{align}
	where $x$ represents the extent to which the state drifts, $z$ represents the amplitude of the convection states, $b_1>0$, $b_2<0$, and $\mu_1$ and $\mu_2$ are two unfolding parameters  that are introduced to respectively represent the deviations ${Pr-Pr^*}$ and ${Ra-Ra^*}$.
	
\begin{figure}
	\centering
	\includegraphics[]{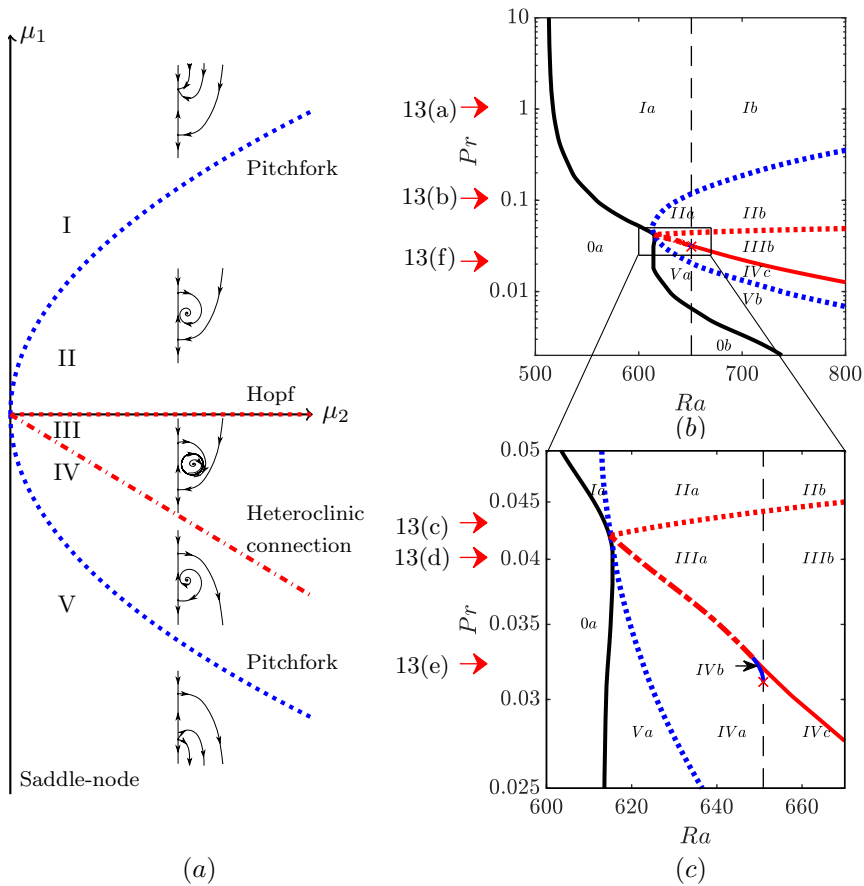}
		\caption{(a) Unfolding near the codimension two saddle-node-pitchfork bifurcation at $\mu_1=\mu_2=0$ given by system (\ref{eq:unfoldingx}, \ref{eq:unfoldingz}), after \cite{guckenheimer1983nonlinear}. 
			The different phase portraits are classified in five different regions labelled using Roman numerals and accompanied with a sketch of the corresponding phase space. In each of these sketches, the fixed points on the vertical line represent steady convection states. 
			The vertical (resp. horizontal) direction is the eigendirection related to the amplitude (resp. drift) mode. 
			(b) Analogy with the doubly diffusive convection problem is made by replacing $\mu_1$ by $Pr-Pr^*$ and $\mu_2$ by $Ra-Ra^*$ and regions of the $(Ra,Pr)$ parameter space are shown as a function of the observed temporal behaviour for $Le = 11$. 
			(c) Magnification of panel (b) near $(Ra^*,Pr^*)$. 
			Arrows indicate the values of $Pr$ used to produce the bifurcation diagrams in figure~\ref{fig:unfoldingbd}. 
			In the panels, the bifurcations are represented by: black and red solid lines (saddle-nodes), blue dotted lines (drift bifurcation), red dotted lines (Hopf bifurcation), red dot-dashed lines (heteroclinic connection) and in panels (b) and (c), the vertical dashed lines (primary stationary bifurcation of the conduction state). }
		\label{fig:unfolding}
	\end{figure}

	When $\mu_1 = \mu_2 = 0$, the trivial state, $(x,z) = (0,0)$, undergoes a codimension two bifurcation.
	One of five phase portraits is observed in the vicinity of this bifurcation and these are shown in figure~\ref{fig:unfolding}(a).
	In addition to the steady states previously discussed, the unfolding reveals the presence of periodic orbits.
	Relating the unfolding back to doubly diffusive convection, these correspond to relative periodic orbits consisting of drifting states that originate either from a travelling wave undergoing a Hopf bifurcation or from a global bifurcation where two steady convection states connect heteroclinically. 
	
\begin{figure}
		\centering
		\includegraphics[width=\linewidth]{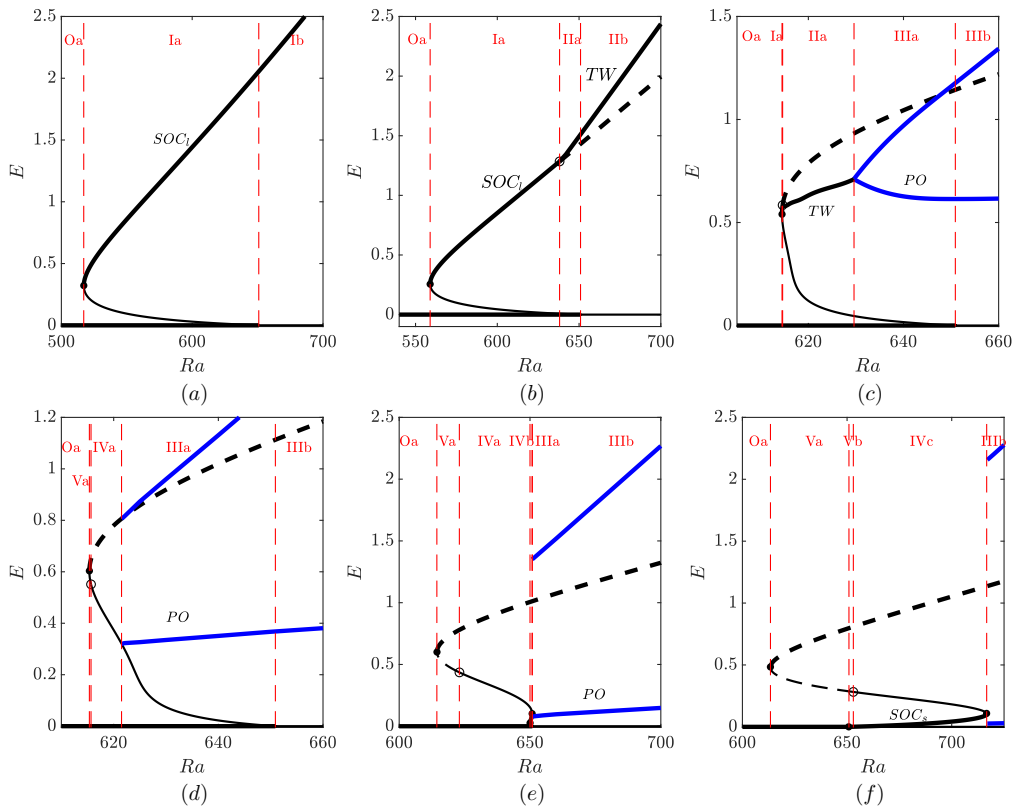}
		\caption{Bifurcation diagrams showing the primary branch and other stable attracting states for $Le = 11$, and (a) $Pr =1$, (b) $Pr = 0.1$, (c) $Pr = 0.043$, (d) $Pr = 0.04$, (e)~$Pr = 0.032$ and (f) $Pr = 0.02$. The solid circles mark the saddle-nodes and open circles indicate where the drift bifurcation occurs. Thick (thin) lines represent states stable (unstable) to the amplitude mode whilst solid (dashed) lines show those stable (unstable) to the drift mode. Thick blue lines indicate the minimal and maximal energies achieved in the stable limit cycle, which starts in a Hopf bifurcation in (c) and in a heteroclinic bifurcation in (d,e,f). The unstable branches of travelling waves are not shown.}
		\label{fig:unfoldingbd}
	\end{figure}

	Although the normal form (\ref{eq:unfoldingx}, \ref{eq:unfoldingz}) only formally represents the dynamics of the full system close to the codimension two point, each of the regions shown in figure~\ref{fig:unfolding} continues to be observed an appreciable distance away from this point.
	Figures~\ref{fig:unfolding}(b) and (c) illustrate the extent of the corresponding regions in the doubly diffusive system when $Le = 11$ and we anticipate that similar results will hold for other values of the Lewis number.
	In this figure, the regions have been subdivided according to the types of stable attracting states that they display.
	The subdivisions occur owing to the instability of the conduction state at $Ra_c$ and the creation of a pair of saddle-nodes at $(Ra_{cusp},Pr_{cusp})$, which enrich the previous unfolding. 
	The resulting subregions, together with their associated attracting states, are summarised in table~\ref{tab:stableattractorregions} and on the bifurcation diagrams in figure~\ref{fig:unfoldingbd}.
	\begin{table}
		\centering
		\begin{tabular}{c|ccccccccccccc}
			\toprule
			& \multicolumn{13}{c}{Stable in region?}  \\
			State & Oa & Ob & Ia& Ib&IIa &IIb &IIIa & IIIb &IVa &IVb &IVc &Va &Vb \\ 
			\midrule
			$O$	&x& &x& &x& &x& &x&x & &x & \\
			$SOC_s$ & &x& & & & & & & &x&x& &x \\
			$SOC_l$ &  & &x &x & & &  & & & & & & \\
			$TW$ &	& & & & x& x& & && & & & \\
			$PO$ &	& & & & & &x &x &&& & & \\
			\bottomrule
		\end{tabular} 
		\caption{Stability of the observed doubly diffusive states within each region of the parameter space from figure~\ref{fig:unfolding}. The naming convention used is as follows: $O$, conduction state; $SOC_s$, small-amplitude stationary overturning convection; $SOC_l$, large-amplitude stationary overturning convection; $TW$, travelling wave; and $PO$, relative periodic orbit. The regions Oa, ..., Vb refer to the regions introduced in figure~\ref{fig:unfolding}. }
		\label{tab:stableattractorregions}
	\end{table}
	As $Pr$ varies, the system admits one of seven qualitatively distinct bifurcation diagrams. 
	Six of these are presented in figure~\ref{fig:unfoldingbd}, which also indicate the range of kinetic energies over each relative periodic orbit attained via time-stepping. 
	The seventh type of bifurcation diagram, where the primary branch lies entirely within the supercritical regime, is not shown but possesses similar features to that seen for $Pr = 0.02$ in figure~\ref{fig:unfoldingbd}(f) including stable small-amplitude steady convection states and relative periodic orbits.
	
\begin{figure}
		\centering
		\includegraphics[width=\linewidth]{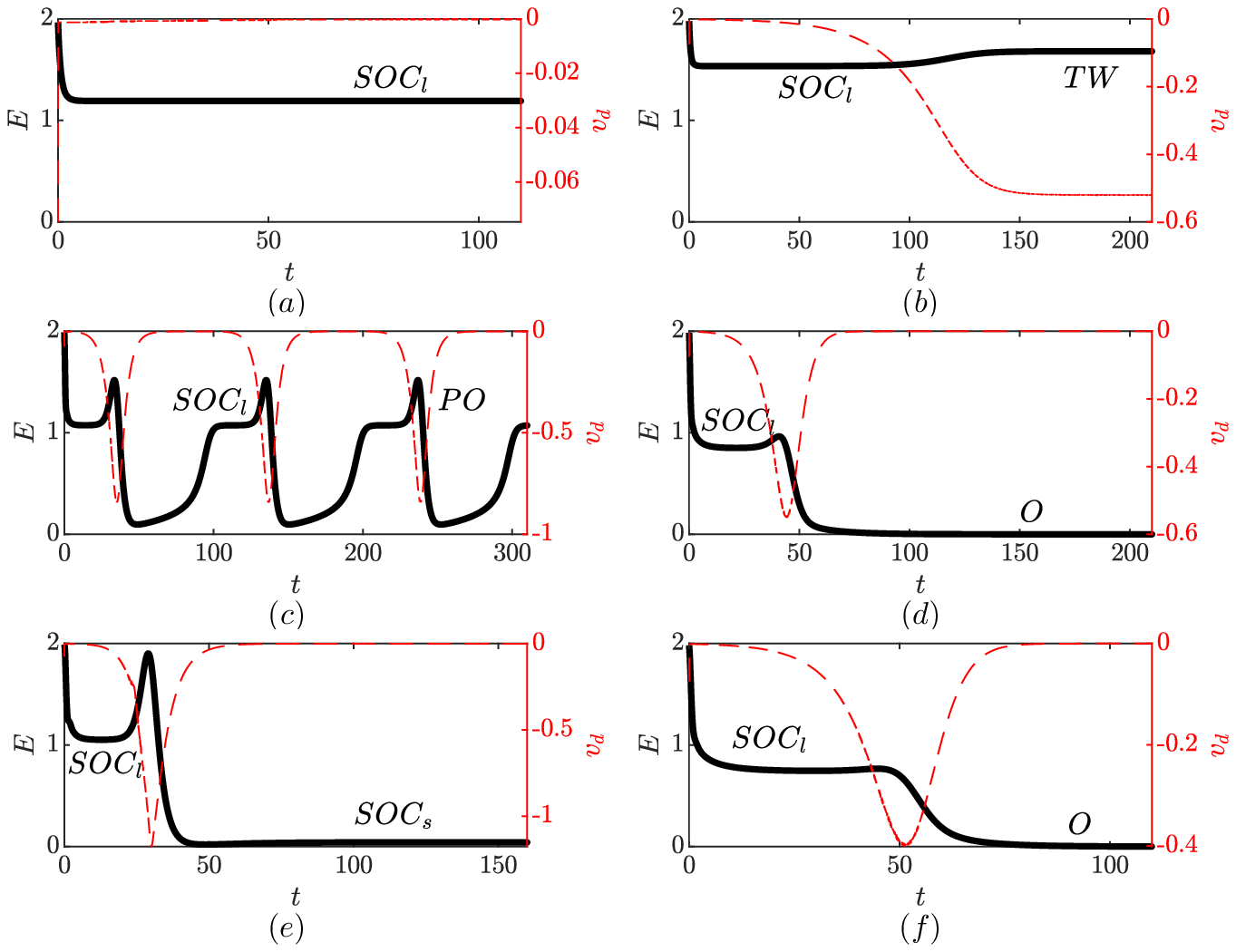}
		\caption{Energy-time (black) and drift speed-time (red) plots illustrating regions I\textendash{}V in figure~\ref{fig:unfolding} with $Le = 11$. In each case, the initial state was the large-amplitude convection state at $Ra = 700$ for $Pr = 0.1$ that was perturbed in the direction of its unstable drift eigenmode. States approached during the trajectory are labelled as follows: (a) region Ia, convergence to $SOC_l$ when $Pr = 0.1$ and $Ra = 630$; (b) region IIb, convergence to $TW$ when $Pr = 0.1$ and $Ra = 660$; (c) region IIIb, convergence to $PO$ when $Pr = 0.032$ and $Ra = 660$, (d) region IVa, convergence to $O$ when $Pr = 0.032$ and $Ra = 630$; (e) region IVc, convergence to $SOC_s$ when $Pr = 0.02$ and $Ra = 700$; and (f) region Va, convergence to $O$ when $Pr = 0.032$ and $Ra = 620$. }
		\label{fig:unfoldingEtplots}
	\end{figure}

	The three most relevant stable attracting states close to the primary bifurcation at high $Pr$ ($Pr>Pr^*$ here) are: the conduction state ($O$), the large-amplitude steady convection states ($SOC_l$) and the travelling wave states ($TW$).
	Below the onset of convection (region Oa), all initial conditions decay towards the first of these.
	In region~Ia, above subcritical onset but before the drift instability, initial conditions converge towards $SOC_l$, as evidenced by the energy-time and drift speed-time plot in figure~\ref{fig:unfoldingEtplots}(a).
	Increasing $Ra$ beyond the drift instability into region IIa, $SOC_l$ is now unstable and the flow converges towards $TW$.
	Figure~\ref{fig:unfoldingEtplots}(b) shows that the former state may still be observed in the temporal dynamics as the initial condition first rapidly changes amplitude to approach $SOC_l$ before it builds vertical drift and converges to $TW$.
	%
	
\begin{figure}
		\centering
		\includegraphics[width=\linewidth]{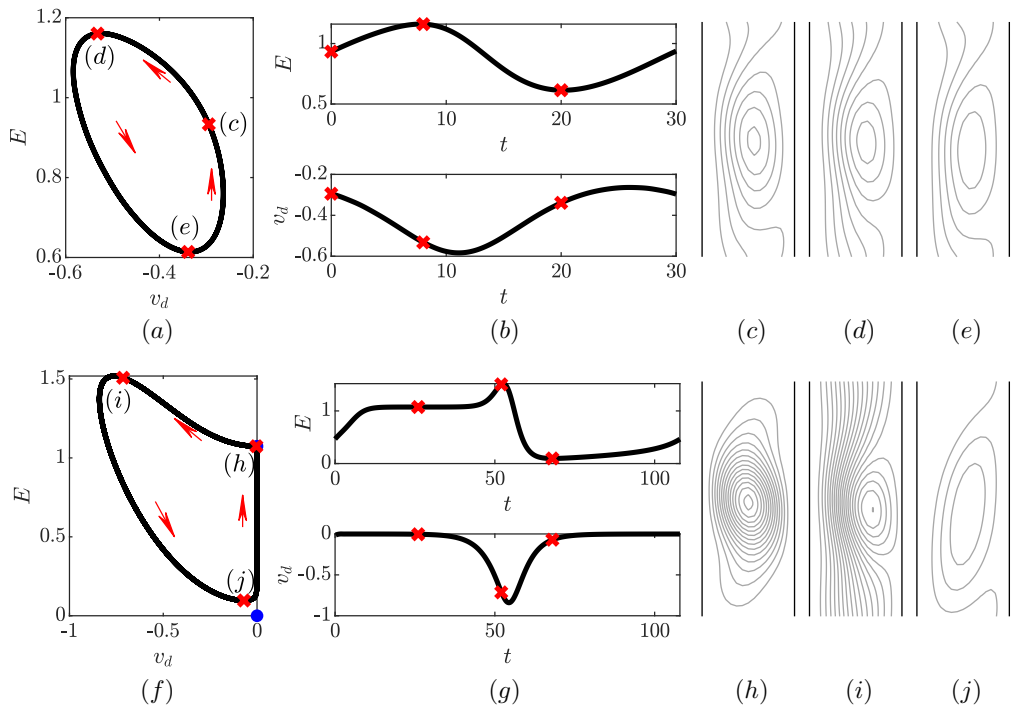}
		\caption{Temporal evolution of downward-travelling states across one cycle of two relative periodic orbits at: (a)\textendash{}(e) $Ra = 650$ with $Pr = 0.043$ and $Le = 11$ and (f)\textendash{}(j) $Ra = 660$ with $Pr = 0.032$ and $Le = 11$ (as in figure~\ref{fig:unfoldingEtplots}(c)). 
			(a) and (f) Anticlockwise trajectory of the periodic orbit in drift speed-energy phase space. Blue dots in (a) mark the conduction and steady convection states. 
			(b) and (g)  Energy-time (top) and drift speed-time (bottom) plots. (c)\textendash{}(e) Streamfunctions of states along the orbit in (a) at (b) $t = 0$, (c) $t = 8$ and (d) $t=20$ with contour intervals $0.1$.
			(h)\textendash{}(i) Streamfunctions of states along the orbit in (f) at (h) $t = 26$, (i) $t = 52$ and (j) $t=68$, with contour intervals $0.05$. 
			The streamfunctions have been translated vertically for better visual representation.}
		\label{fig:ts660le11pr0032}
	\end{figure}

	The stable branch of travelling waves destabilises in a supercritical Hopf bifurcation that leads to a stable relative periodic orbit, as shown in figure~\ref{fig:unfoldingbd} for $Pr = 0.043$.
	Figures~\ref{fig:ts660le11pr0032}(a)\textendash{}(e) depict such an orbit shortly after the bifurcation at $Ra = 650$ and $Pr = 0.043$, where we see that the states exhibit small oscillations about a drifting state.
	The Hopf bifurcation moves towards lower Rayleigh numbers as $Pr$ approaches $Pr^*$ from above, which reduces the extent over which stable $TW$ are found.
	This continues until $Pr = Pr^*$, when stable $TW$ cease to exist and the relative periodic orbit bifurcates directly from the codimension two bifurcation at the saddle-node. 
	
	Upon further decreasing of the Prandtl number, so that the drift bifurcation occurs on the lower branch of steady convection, the system admits neither stable $SOC_l$ nor stable $TW$.
	Instead, the bifurcation diagrams are similar to that shown for $Pr = 0.04$ in figure~\ref{fig:unfoldingbd}(d), where a branch of unstable $TW$ extends from the drift bifurcation towards higher Rayleigh numbers and stable relative periodic orbits exist after a global bifurcation, where the stable manifold of $SOC_l$ connects heteroclinically with the unstable manifold of the convection state on the lower branch and vice versa.
	
	The lack of stability of the nonlinear states before the heteroclinic connection lead all initial conditions to decay down to the conduction state in regions IVa and Va. 
	Figures~\ref{fig:unfoldingEtplots}(d) and (f) illustrate this tendency for $Pr = 0.032$ when $Ra = 630$ and $Ra = 620$, respectively.
	In both cases, the amplitude of the initially imposed roll rapidly decreases to approach that of $SOC_l$ rolls.
	Afterwards, the drift speed of the state increases, as $SOC_l$ is unstable to drift, and reaches a maximum around $t \approx 50$.
	The drift speed subsequently decays down to zero, due to the instability of $TW$, and the time-dependent state converges on the conduction state, which is the only stable attractor in these regions.
	
	Beyond the heteroclinic connection, initial conditions tend to converge towards the relative periodic orbit, as they invariably do in region IIIb, where the conduction state is unstable.
	Figure~\ref{fig:unfoldingEtplots}(c) illustrates this convergence starting from a large-amplitude roll with $Pr = 0.032$ and $Ra = 660$ perturbed in the direction of its unstable drift eigenmode.
	A single cycle of this orbit is shown in further detail in figures~\ref{fig:ts660le11pr0032}(f)\textendash{}(j).
	This relative periodic orbit cycles between the three states: $SOC_l$, $TW$ and a steady small-amplitude convection state, in the following manner.
	The first stage of the orbit, from $15\lesssim t \lesssim 40$, resembles the temporal behaviour seen in region IIa (figure~\ref{fig:unfoldingEtplots}(b)), where the solution remains close to $SOC_l$ in profile (figure~\ref{fig:ts660le11pr0032}(h)) while the drift speed slowly increases in magnitude.
	Following this, between $t\approx 40$ and $t\approx 54$, the drift speed and kinetic energy rapidly increase as the profile of the state exhibits properties of the travelling wave ($TW$) solution (figure~\ref{fig:ts660le11pr0032}(i)).
	Between $t\approx 54$ and $t \approx 68$, both the drift speed and kinetic energy decrease as the state approaches a small-amplitude, non-drifting convection state with inclined rolls (figure~\ref{fig:ts660le11pr0032}(j)).
	The final stage of this orbit is the transition from the small-amplitude back to large-amplitude steady convection, which is indicated by the monotonic increase in kinetic energy while maintaining $v_d \approx 0$ for $t \gtrsim 70$ and $t \lesssim 15$ in figure~\ref{fig:ts660le11pr0032}(g).
	
\begin{figure}
		\centering
		\includegraphics[width=\linewidth]{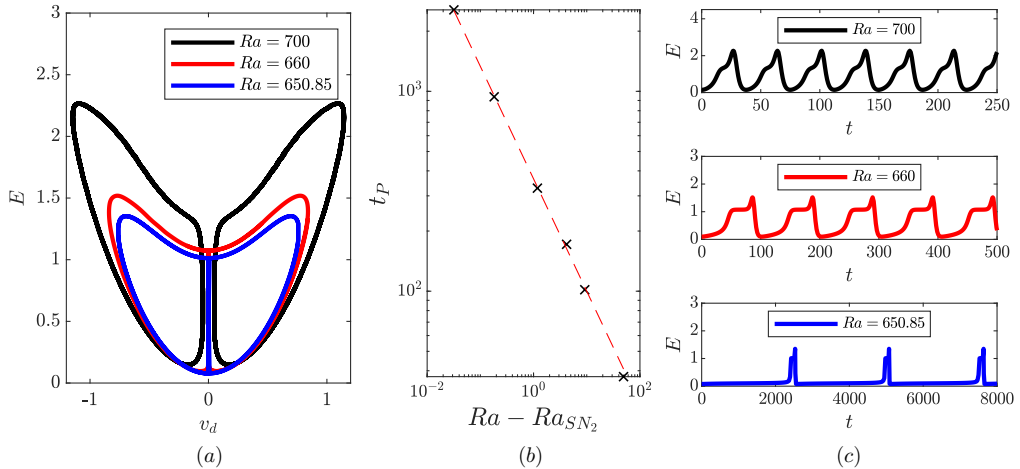}
		\caption{Relative periodic orbits for $Le = 11$, $Pr = 0.032$ where $Ra_{SN_2} \approx 650.82$. (a) Trajectories in $(v_d,E)$ phase space for $Ra = 650.85$ (blue), $Ra=660$ (red) and $Ra=700$ (black).
		For $Ra = 700$ and $Ra = 660$, a pair of relative periodic orbits associated with either negative or positive drift velocity are shown, while for $Ra = 650.85$, a single periodic orbit with alternating negative and positive drift velocities is shown.
		 (b) Period $t_P$ of orbits for selected $Ra>Ra_{SN_2}$. The red dashed line shows that approximately $t_P \propto (Ra-Ra_{SN_2})^{-0.56}$. (c)~Energy-time plots for $Ra = 700$ (top), $Ra = 660$ (middle) and $Ra = 650.85$ (bottom).}
		\label{fig:pophasespacethreele11pr0032}
	\end{figure}

	The heteroclinic connection leading to these orbits moves towards higher Rayleigh numbers as $Pr$ decreases and  coincides with $SN_2$ for $Pr \lesssim 0.032$ (see figures~\ref{fig:unfoldingbd}(e) and (f)).
	This suggests that a saddle-node infinite period (SNIPER) bifurcation explains the origin of the relative periodic orbits at low Prandtl and high Rayleigh numbers.
	However, by considering various properties of the relative periodic orbits for $Pr = 0.032$ and $Le = 11$ as $Ra$ approaches $Ra_{SN_2}$ from above (figure~\ref{fig:pophasespacethreele11pr0032}), we additionally find that a gluing bifurcation occurs in the vicinity of the SNIPER bifurcation.
	
	At large Rayleigh numbers, a pair of relative periodic orbits with states drifting either upwards or downwards 
	are related by the reflection symmetry.
	The maximal energy and drift speed attained along these orbits decrease with decreasing Rayleigh number, and the trajectories approach the stable and unstable manifolds of $SOC_l$, as seen in figure~\ref{fig:pophasespacethreele11pr0032}(a).
	This leads to the two relative periodic orbits connecting in a gluing bifurcation around $Ra \approx 652$ so that the trajectories become a single periodic orbits where states alternately drift in opposite directions.
	This is reminiscent of the pulsating waves seen in nonlinear magnetoconvection \citep{matthews1993pulsating}.

	The resulting single periodic orbit persists until $Ra_{SN_2}$, where it terminates in the SNIPER bifurcation.
	This is evidenced by the period of a single loop of the orbit scaling like $t_P \propto (Ra-Ra_{SN_2})^{-0.56}$ as $SN_2$ is approached, which is close to the expected $t_P \approx |Ra-Ra_{SN_2}|^{-0.5}$ scaling. 	
	The energy-time plots in figure~\ref{fig:pophasespacethreele11pr0032}(c) illustrate that the predominant increase in duration occurs near the small-amplitude steady convection state as the orbit approaches the steady state at $SN_2$ in phase space.
	We also find that the time spent near $SOC_l$ increases, whilst the time where the state has large drift speed remains small, implying that the global bifurcation is due to the collision of the periodic orbit with the stable manifold of $SOC_l$.
	
	The final attracting state that the flow may converge to is $SOC_s$, as figure~\ref{fig:unfoldingEtplots}(e) illustrates for $Pr = 0.02$ and $Ra = 700$.
	This is possible for $Pr < Pr_{cusp}$ in the supercritical regions Ob, IVc and Vb, where it is the only stable attracting state, and in the subcritical region IVb, where convergence towards the stable conduction state is also possible.

	\section{Discussion}\label{sec:discussion}
	This paper considers doubly diffusive convection driven by horizontal gradients of temperature and concentration, a configuration typically referred to as natural doubly diffusive convection.
	We have extended the linear stability analysis of \citet{ghorayeb1997double} by performing a thorough weakly nonlinear analysis of the system.
	This was complemented by a numerical exploration of the nonlinear regime, thereby also extending the analysis of \citet{xin1998bifurcation}, who focused on $Pr = 1$ and $Le = 1.2$.
	From this analysis, we unravelled the relationships between saddle-nodes, drift and global bifurcations. 
	
	We have identified regions where the resulting primary branch exhibits qualitatively different behaviour.
	For large values of the Prandtl number, the bifurcation is subcritical and hysteresis takes place between the conduction state and large-amplitude convection.
	Whereas, for Prandtl numbers below a critical value, the primary bifurcation is supercritical but this is preceded by the creation of two saddle-nodes without affecting the existence of large-amplitude convection. 
	Despite this, we did not find any hysteresis in the supercritical regime owing to the presence of a destabilising drift bifurcation along the primary branch.
	The presence of multiple folds along a primary supercritical branch has already been observed in a non-homogeneous fluid system \citep{erenburg2003multiple} but we believe that is the first time that it has been observed in homogeneously forced convection in such a small domain. 
	
	By determining the stability of steady convection states along the primary branch, we identified a codimension two point between a large-amplitude saddle-node and a drift bifurcation.
	We analysed the dynamics around this codimension two point using its normal form and numerical simulations to investigate new Hopf and heteroclinic bifurcations giving rise to periodic orbits. 
	Such time-dependent states are common features of low Prandtl number doubly diffusive convection (see also \citet{umbria2019stationary}).
	Finally, we provided a classification of the various regions in $(Ra,Pr)$ parameter space according to the nature of their dynamical attractors, for a representative value of the Lewis number.
	
	We anticipate that the analysis provided in this paper may serve as a guide for future research in natural doubly diffusive convection by providing a comprehensive map of the near-onset dynamics as a function of the parameter values. 
	Despite our attempt to be thorough, the characterisation of the nonlinear regime at very small Prandtl numbers, which is relevant in astrophysical contexts (see \citet{garaud2018}), remains to be explored.
	We observed that this regime behaves differently from extrapolated predictions from $O(1)$ Prandtl numbers but have not pursued this any further.

	Lastly, the coexistence of steady overturning convection with the stable conduction state when the primary bifurcation is supercritical has important dynamical implications, which will be the subject of future exploration.
	In particular, it makes this system a candidate for spatially localised pattern formation in a supercritical fluid system, owing to the similarity of the primary branch structure with the Swift--Hohenberg equation considered by \citet{knobloch2019defectlike}.
	
	\bigskip
	
	\noindent {\bf Acknowledgements}
	
	This work was supported by the Leeds\textendash{}York Natural Environment Research Council (NERC) Doctoral Training Partnership (DTP) SPHERES under grant NE/L002574/1 and undertaken on ARC3, part of the High Performance Computing facilities at the University of Leeds, UK.\\
        
	\noindent {\bf Declaration of Interests}

        The authors report no conflict of interest.
        
	\appendix
	
	\section{Further expressions for the weakly nonlinear analysis}\label{sec:abgd_exp}
	
	\subsection{Second-order corrections}\label{sec:O2corrections}
	The solution to the system at $\mathcal{O}(\epsilon^2)$ (\ref{eq:O(eps^2)}) given in (\ref{eq:dep_2}) involves parameter-free functions $\tilde{u}_i, \tilde{w}_i, \tilde{p}_i$ and $\tilde{\theta}_i$ for $i =2,...,7$ within the expressions for $\Psi_2^0, \Psi_2^1$ and $\Psi_2^2$ (\ref{eq:dep_psi_0}\textendash\ref{eq:dep_psi_2}).
	These functions satisfy the forced linear equations:
	\begin{align}
	\begin{pmatrix}
	-D & 0 & 0 & 0 \\
	0 & D^2 & 0 & 0\\
	0 & 0 & D^2 &  Ra_c(1-Le) \\
	0 & 0 & 0 & D^2\\ 
	\end{pmatrix}
	\begin{pmatrix}
	\tilde{p}_2 \\ \tilde{w}_2\\\tilde{w}_3 \\ \tilde{\theta}_3
	\end{pmatrix}&=
	\begin{pmatrix}
	f_{10}\\f_{20}\\0 \\f_{40}
	\end{pmatrix}, \label{eq:O2_0}\\
	\begin{pmatrix}
	D^2-4k_c^2 & 0 & -D & 0 \\
	0 & D^2 -4k_c^2 & -2ik_c & Ra_c(1-Le) \\
	D & 2ik_c & 0 & 0\\
	-1 & 0 & 0 & D^2-4k_c^2\\
	\end{pmatrix} 
	\begin{pmatrix}
	\tilde{u}_4\\\tilde{w}_4\\\tilde{p}_4\\\tilde{\theta}_4\\
	\end{pmatrix}
	&= 
	\begin{pmatrix}
	f_{12}\\f_{22}\\0\\ 0
	\end{pmatrix},\label{eq:O2_Pr}\\
	\begin{pmatrix}
	D^2-4k_c^2 & 0 & -D & 0& 0\\
	0 & D^2 -4k_c^2 & -2ik_c & Ra_c(1-Le) & 0\\
	D & 2ik_c & 0 & 0 &0\\
	-1 & 0 & 0 &D^2 -4k_c^2 & 0\\
	-1 & 0 & 0 & 0 &D^2-4k_c^2\\
	\end{pmatrix}
	\begin{pmatrix}
	\tilde{u}_5\\\tilde{w}_5\\\tilde{p}_5\\\tilde{\theta}_5 \\\tilde{\theta}_6
	\end{pmatrix} &=
	\begin{pmatrix}
	0 \\ 0\\ 0 \\f_{42} \\0
	\end{pmatrix}, \label{eq:O2_1Le}\\
	\begin{pmatrix}
	D^2-k_c^2 & 0 & -D & 0\\
	0 & D^2 -k_c^2 & -ik_c & Ra_c(1-Le) \\
	D & ik_c & 0 & 0 \\
	-1 & 0 & 0 &D^2 -k_c^2 
	\end{pmatrix}
	\begin{pmatrix}
	\tilde{u}_7\\\tilde{w}_7\\\tilde{p}_7\\\tilde{\theta}_7 \\
	\end{pmatrix} &=
	\begin{pmatrix}
	f_{11}\\ f_{21}\\ f_{31} \\f_{41} 
	\end{pmatrix}, \label{eq:O2_1}
	\end{align}
	where $D = \frac{d}{dx}$, and $\tilde{u}_i$, $\tilde{w}_i$ and $\tilde{\theta}_i$ satisfy homogeneous boundary conditions and the pressure boundary conditions come from a projection of the Navier\textendash{}Stokes equation onto the side walls:
	\begin{align}
	\tilde{u}_i = 0,\quad \tilde{w}_i = 0, \quad -\frac{\partial \tilde{p}_i}{\partial x} + \frac{\partial^2 \tilde{u}_i}{\partial x^2}=0 \quad \tilde{\theta}_i = 0 \quad \text{ on } \quad x = 0,1 \label{eq:bc0O2}
	\end{align}
	\subsection{Coefficients in the Ginzburg\textendash{}Landau equation}\label{sec:GLEcoeff}
	Expressions for the coefficients $\alpha$, $\beta$, $\gamma$ and $\delta$ in the Ginzburg\textendash{}Landau equation (\ref{eq:GLE1}) are obtained by evaluating:
	\begin{align}
	\alpha &= \frac{1}{Pr}\left(\langle U^{\dagger}\,,\,U_1\rangle + \langle W^{\dagger}\,,\,W_1\rangle\right) + (1 + Le)\langle \Theta^{\dagger}\,,\,\Theta_1\rangle\notag\\
	&= \frac{1}{Pr}\alpha_1 + (1+Le)\alpha_2,\label{eq:alp2}\\
	\beta &= -\left(\frac{1}{Pr}\langle U^{\dagger}\,,\,\mathcal{N}_3^U\rangle + \frac{1}{Pr}\langle W^{\dagger}\,,\,\mathcal{N}_3^W\rangle 
	+ \frac{1}{1-Le}\langle \Theta^{\dagger}\,,\,\mathcal{N}_3^\Theta-Le\mathcal{N}_3^{\Phi}\rangle 
	\right)\notag
	\\&=\frac{1}{Pr^2}\beta_1 + \frac{1+Le}{Pr}\beta_2 + (1+Le^2)\beta_3+Le\beta_4, \label{eq:bet2}\\
	\gamma &= (1-Le)\langle W^{\dagger},\Theta_1\rangle \notag\\
	&= (1-Le)\gamma_1,\label{eq:gam2}\\
	\begin{split}\delta &=\langle U^{\dagger}\,,\,U_1\rangle + \langle W^{\dagger}\,,\,W_1\rangle + \langle \Theta^{\dagger}\,,\,\Theta_1\rangle\\
	&\quad {}+2ik_c\left(\langle U^{\dagger}\,,\,\tilde{u}_7\rangle + \langle W^{\dagger}\,,\,\tilde{w}_7\rangle + \langle \Theta^{\dagger}\,,\,\tilde{\theta}_7\rangle\right) + \langle P^{\dagger}\,,\,\tilde{w}_7\rangle -\langle W^{\dagger}\,,\,\tilde{p}_7\rangle, \end{split} \label{eq:del}
	\end{align}
	where the nonlinear functions $\mathcal{N}_3^F$, for $F = U,W,\Theta,\Phi$, and coefficients $\beta_i$ for $i = 1,2,3,4$ are:
	\begin{align}
	\mathcal{N}_3^F &= U_1\frac{dF_2^0}{dx}+\overline{U}_1\frac{dF_2^2}{dx} + U_2^2\frac{d\overline{F}_1}{dx}+ 2ik_c\overline{W}_1F_2^2  + ik_cW_2^0F_1 -ik_cW_2^2\overline{F}_1,\\
	\begin{split}\beta_1 &= 
	{}-\left\langle U^{\dagger}, \overline{U}_1\frac{d\tilde{u}_4}{dx} + \tilde{u}_4\frac{d\overline{U}_1}{dx} +2ik_c\tilde{u}_4\overline{W}_1 +ik_c\tilde{w}_2U_1 -ik_c\tilde{w}_4\overline{U}_1\right\rangle \\
	&\quad\;- \left\langle W^{\dagger}, U_1\frac{d\tilde{w}_2}{dx} + \overline{U}_1\frac{d\tilde{w}_4}{dx} + \tilde{u}_4\frac{d\overline{W}_1}{dx} + ik_c\overline{W}_1\tilde{w}_4 +ik_cW_1\tilde{w}_2 \right\rangle,\end{split} \label{eq:a_beta1}\\
	\begin{split} \beta_2 &= {}-\left\langle U^{\dagger},\overline{U}_1\frac{d\tilde{u}_5}{dx} + \tilde{u}_5\frac{d\overline{U}_1}{dx} +2ik_c\tilde{u}_5\overline{W}_1 +ik_c\tilde{w}_3U_1 -ik_c\tilde{w}_5\overline{U}_1 \right\rangle \\
	&\quad\;{}-\left\langle W^{\dagger},U_1\frac{d\tilde{w}_3}{dx} + \overline{U}_1\frac{d\tilde{w}_5}{dx} + \tilde{u}_5\frac{d\overline{W}_1}{dx} + ik_c\overline{W}_1\tilde{w}_5 + ik_cW_1\tilde{w}_3\right\rangle\\
	&\quad\;{}-\left\langle \Theta^{\dagger},\overline{U}_1\frac{d\tilde{\theta}_4}{dx} + \tilde{u}_4\frac{d\overline{\Theta}_1}{dx} +2ik_c\overline{W}_1\tilde{\theta}_4 +ik_c\tilde{w}_2\Theta_1 -ik_c\tilde{w}_4\overline{\Theta}_1\right\rangle,
	\end{split}\label{eq:a_beta2}\\
	\begin{split} \beta_3 &= {}-\left\langle \Theta^{\dagger},U_1\frac{d\tilde{\theta}_3}{dx} + \overline{U}_1\frac{d\tilde{\theta}_5}{dx} + \tilde{u}_5\frac{d\overline{\Theta}_1}{dx} +2ik_c\overline{W}_1\tilde{\theta}_5 + ik_c\tilde{w}_3\Theta_1 -ik_c\tilde{w}_5\overline{\Theta}_1\right\rangle,
	\end{split} \label{eq:a_beta3}\\
	\begin{split} \beta_4 &= {}-\left\langle \Theta^{\dagger},U_1\frac{d\tilde{\theta}_3}{dx} + \overline{U}_1\left(\frac{d\tilde{\theta}_5}{dx} +\frac{d\tilde{\theta}_6}{dx}\right)+ 2\tilde{u}_5\frac{d\overline{\Theta}_1}{dx}      \right.\qquad \qquad \qquad \qquad\\
	&{}\qquad \qquad \qquad\left.+2ik_c\left(\overline{W}_1\left(\tilde{\theta}_5 +\tilde{\theta}_6\right)+ \tilde{w}_3\Theta_1 -\tilde{w}_5\overline{\Theta}_1\right)\right\rangle
	\end{split}. \label{eq:a_beta4}
	\end{align}
	These expressions for the parameter-free coefficients $\alpha_i$, $\beta_i$, $\gamma_1$ and $\delta$ are evaluated numerically and are given in table~\ref{tab:coeff_values}.
	
	\begin{table*}
		\centering
		\begin{tabular}{rrrrrrrr}
			\toprule
			$\alpha_1$ &$1.11\times10^{-4}$ &\hspace{0.25cm} &$\beta_2$ & $-1.63\times10^{-8}$&\hspace{0.25cm} &$(Le > 1)$ $\gamma_1$ & $-8.85\times10^{-7}$\\
			$\alpha_2$ & $2.27\times10^{-4}$&&$\beta_3$ &$7.47\times10^{-8}$ & &$(Le < 1)$ $\gamma_1$  & $8.85\times10^{-7}$\\
			$\beta_1$ & $-4.43\times10^{-9}$&&$\beta_4$ &$1.58\times10^{-7}$ & &$\delta$ & $7.38\times10^{-4}$ \\
			\bottomrule
		\end{tabular} 
		\caption{Numerical values of the coefficients $\alpha_1$, $\alpha_2$, $\beta_1$, $\beta_2$, $\beta_3$, $\beta_4$, $\gamma_1$ and $\delta$ in (\ref{eq:GLE1}). The sign of $\gamma_1$ depends upon whether $Le >1$ or $Le < 1$ as $\gamma>0$ for all $Le$, while all other coefficients are independent of the parameters $Le$ and $Pr$.}
		\label{tab:coeff_values}
	\end{table*}
	
	Of particular interest is the boundary where the primary bifurcation changes from subcritical to supercritical. This occurs when $\beta = 0$, which we may find explicitly by taking the positive root of equation (\ref{eq:bet2}), to find:
	\begin{equation}
	Pr_c = \frac{-(1+Le)\beta_2 + \sqrt{(1+Le)^2\beta_2^2-4\beta_1\left[(1+Le^2)\beta_3+Le\beta_4\right]}}{2[(1+Le^2)\beta_3+Le\beta_4]}.\label{eq:Prc}
	\end{equation}
	
	\subsection{Effect of thermal and solution advective terms on $a_2$}\label{sec:adv}
\begin{figure}
		\centering
		\includegraphics[]{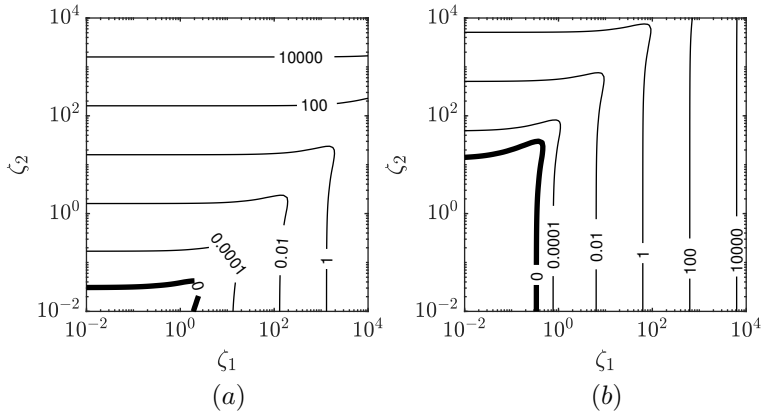}
		\caption{Contours of the coefficient $a_2$ as a function of $\zeta_1$ and  $\zeta_2$, which respectively multiply thermal and solutal advective nonlinearities in (\ref{eq:zetaNS}\textendash{}\ref{eq:zetaC}), for (a) $Le = 11$, $Pr = 1$ and (b) $Le = 1/11$, $Pr = 1$.
			The contour $a_2 = 0$, which marks the boundary between subcriticality and supercriticality, is shown in bold.}
		\label{fig:zeta2le09}
	\end{figure}

	To determine the contributions that each of the nonlinear terms make to $a_2$, we introduce the factors $\zeta_1$ and $\zeta_2$ that multiply the thermal and solutal advective terms, respectively.
	We numerically perform the weakly nonlinear analysis for the modified system:
	\begin{align}
	\frac{1}{Pr}\left(\frac{\partial \mathbf{u}}{\partial t} + \mathbf{u}\cdot \nabla \mathbf{u}\right)&= - \nabla p + \nabla^2 \mathbf{u} + Ra(T-C)\hat{\mathbf{z}}, \label{eq:zetaNS}\\
	\nabla \cdot \mathbf{u} &= 0,\\
	\frac{\partial T}{\partial t} + \zeta_1 \mathbf{u}\cdot \nabla T &= \nabla^2 T,\\
	\frac{\partial C}{\partial t} + \zeta_2 \mathbf{u}\cdot \nabla C &= \frac{1}{Le}\nabla^2 C, \label{eq:zetaC}
	\end{align}
	with $\zeta_1$, $\zeta_2 \in [10^{-2},10^4]$ and selected values of the Prandtl and Lewis numbers.
	The coefficient $a_2$ tends to increase when one of $\zeta_1$ or $\zeta_2$ increases, while keeping the other fixed, as indicated by the contours in figure~\ref{fig:zeta2le09}. 
	Thus, temperature and solutal advection enhances the subcriticality of the primary bifurcation.
	
	\bibliographystyle{jfm}
	\bibliography{ref}

\begin{thebibliography}{58}
\expandafter\ifx\csname natexlab\endcsname\relax\def\natexlab#1{#1}\fi
\def\au#1{#1} \def\ed#1{#1} \def\yr#1{#1}\def\at#1{#1}\def\jt#1{\textit{#1}}
  \def\bt#1{#1}\def\bvol#1{\textbf{#1}} \def\vol#1{#1} \def\pg#1{#1}
  \def\publ#1{#1}\def\arxiv#1{#1}\def\org#1{#1}\def\st#1{\textit{#1}}

\bibitem[Batiste {\em et~al.\/}(2001)Batiste, Knobloch, Mercader \&
  Net]{batiste2001}
{\sc \au{Batiste, O.}, \au{Knobloch, E.}, \au{Mercader, I.} \& \au{Net, M.}}
  \yr{2001}  \at{Simulations of oscillatory binary fluid convection in large
  aspect ratio containers}.  \jt{Phys. Rev. E}  \bvol{65},  \pg{016303}.

\bibitem[Beaume(2017)]{beaume2017adaptive}
{\sc \au{Beaume, C.}} \yr{2017}  \at{Adaptive {Stokes} preconditioning for
  steady incompressible flows}.  \jt{Commun. Comput. Phys.}  \bvol{22}~(2),
  \pg{494--516}.

\bibitem[Beaume(2020)]{beaume2020transition}
{\sc \au{Beaume, C.}} \yr{2020}  \at{Transition to doubly diffusive chaos}.
  \jt{Phys. Rev. Fluids}  \bvol{5}~(10),  \pg{103903}.

\bibitem[Beaume {\em et~al.\/}(2011)Beaume, Bergeon \& Knobloch]{beaume2011}
{\sc \au{Beaume, C.}, \au{Bergeon, A.} \& \au{Knobloch, E.}} \yr{2011}
  \at{Homoclinic snaking of localized states in doubly diffusive convection}.
  \jt{Phys. Fluids}  \bvol{23},  \pg{094102}.

\bibitem[Beaume {\em et~al.\/}(2013{\natexlab{{\em a\/}}})Beaume, Bergeon \&
  Knobloch]{beaume2013convectons}
{\sc \au{Beaume, C.}, \au{Bergeon, A.} \& \au{Knobloch, E.}}
  \yr{2013{\natexlab{{\em a\/}}}}  \at{Convectons and secondary snaking in
  three-dimensional natural doubly diffusive convection}.  \jt{Phys. Fluids}
  \bvol{25}~(2),  \pg{024105}.

\bibitem[Beaume {\em et~al.\/}(2013{\natexlab{{\em b\/}}})Beaume, Bergeon \&
  Knobloch]{beaume2013b}
{\sc \au{Beaume, C.}, \au{Bergeon, A.} \& \au{Knobloch, E.}}
  \yr{2013{\natexlab{{\em b\/}}}}  \at{Nonsnaking doubly diffusive convectons
  and the twist instability}.  \jt{Phys. Fluids}  \bvol{25},  \pg{114102}.

\bibitem[Beaume {\em et~al.\/}(2018)Beaume, Bergeon \&
  Knobloch]{beaume2018three}
{\sc \au{Beaume, C.}, \au{Bergeon, A.} \& \au{Knobloch, E.}} \yr{2018}
  \at{Three-dimensional doubly diffusive convectons: instability and transition
  to complex dynamics}.  \jt{J. Fluid Mech.}  \bvol{840},  \pg{74--105}.

\bibitem[Bergeon {\em et~al.\/}(1999)Bergeon, Ghorayeb \& Mojtabi]{bergeon1999}
{\sc \au{Bergeon, A.}, \au{Ghorayeb, K.} \& \au{Mojtabi, A.}} \yr{1999}
  \at{Double diffusive convection in an inclined cavity}.  \jt{Phys. Fluids}
  \bvol{11},  \pg{549--559}.

\bibitem[Bergeon \& Knobloch(2002)]{bergeon2002natural}
{\sc \au{Bergeon, A.} \& \au{Knobloch, E.}} \yr{2002}  \at{Natural doubly
  diffusive convection in three-dimensional enclosures}.  \jt{Phys. Fluids}
  \bvol{14}~(9),  \pg{3233--3250}.

\bibitem[Bergeon \& Knobloch(2008{\natexlab{{\em a\/}}})]{bergeon2008periodic}
{\sc \au{Bergeon, A.} \& \au{Knobloch, E.}} \yr{2008{\natexlab{{\em a\/}}}}
  \at{Periodic and localized states in natural doubly diffusive convection}.
  \jt{Phys. D: Nonlinear Phenom.}  \bvol{237}~(8),  \pg{1139--1150}.

\bibitem[Bergeon \& Knobloch(2008{\natexlab{{\em b\/}}})]{bergeon2008spatially}
{\sc \au{Bergeon, A.} \& \au{Knobloch, E.}} \yr{2008{\natexlab{{\em b\/}}}}
  \at{Spatially localized states in natural doubly diffusive convection}.
  \jt{Phys. Fluids}  \bvol{20}~(3),  \pg{034102}.

\bibitem[Bethe(1990)]{bethe1990}
{\sc \au{Bethe, H.~A.}} \yr{1990}  \at{Supernova mechanisms}.  \jt{Rev. Mod.
  Phys.}  \bvol{62},  \pg{801--866}.

\bibitem[Clever \& Busse(1981)]{clever1981low}
{\sc \au{Clever, R.~M.} \& \au{Busse, F.~H.}} \yr{1981}
  \at{Low-{Prandtl}-number convection in a layer heated from below}.  \jt{J.
  Fluid Mech.}  \bvol{102},  \pg{61--74}.

\bibitem[Cross {\em et~al.\/}(1983)Cross, Daniels, Hohenberg \&
  Siggia]{Cross1983}
{\sc \au{Cross, M.~C.}, \au{Daniels, P.~G.}, \au{Hohenberg, P.~C.} \&
  \au{Siggia, E.~D.}} \yr{1983}  \at{Phase-winding solutions in a finite
  container above the convective threshold}.  \jt{J. Fluid Mech.}  \bvol{127},
  \pg{155--183}.

\bibitem[Deane {\em et~al.\/}(1988)Deane, Knobloch \& Toomre]{deane1988}
{\sc \au{Deane, A.~E.}, \au{Knobloch, E.} \& \au{Toomre, J.}} \yr{1988}
  \at{Traveling waves in large-aspect-ratio thermosolutal convection}.
  \jt{Phys. Rev. A}  \bvol{37},  \pg{1817--1820}.

\bibitem[Dijkstra \& Kranenborg(1996)]{dijkstra1996bifurcation}
{\sc \au{Dijkstra, H.~A.} \& \au{Kranenborg, E.~J.}} \yr{1996}  \at{A
  bifurcation study of double diffusive flows in a laterally heated stably
  stratified liquid layer}.  \jt{Int. J. Heat Mass Transf.}  \bvol{39}~(13),
  \pg{2699--2710}.

\bibitem[Erenburg {\em et~al.\/}(2003)Erenburg, Gelfgat, Kit, Bar-Yoseph \&
  Solan]{erenburg2003multiple}
{\sc \au{Erenburg, V.}, \au{Gelfgat, A.~Y.}, \au{Kit, E.}, \au{Bar-Yoseph,
  P.~Z.} \& \au{Solan, A.}} \yr{2003}  \at{Multiple states, stability and
  bifurcations of natural convection in a rectangular cavity with partially
  heated vertical walls}.  \jt{J. Fluid Mech.}  \bvol{492},  \pg{63--89}.

\bibitem[Garaud(2018)]{garaud2018}
{\sc \au{Garaud, P.}} \yr{2018}  \at{Double-diffusive convection at low
  {Prandtl} number}.  \jt{Annu. Rev. Fluid Mech.}  \bvol{50},  \pg{50}.

\bibitem[Ghorayeb \& Mojtabi(1997)]{ghorayeb1997double}
{\sc \au{Ghorayeb, K.} \& \au{Mojtabi, A.}} \yr{1997}  \at{Double diffusive
  convection in a vertical rectangular cavity}.  \jt{Phys. Fluids}
  \bvol{9}~(8),  \pg{2339--2348}.

\bibitem[Guckenheimer \& Holmes(1983)]{guckenheimer1983nonlinear}
{\sc \au{Guckenheimer, J.} \& \au{Holmes, P.}} \yr{1983} {\em Nonlinear
  Oscillations, Dynamical Systems and Bifurcations of Vector Fields\/}.
  \publ{New York: Springer}.

\bibitem[Huppert \& Sparks(1984)]{huppert1984}
{\sc \au{Huppert, H.~E.} \& \au{Sparks, R. S.~J.}} \yr{1984}  \at{Double
  diffusive convection due to crystallization in magmas}.  \jt{Annu. Rev. Earth
  Planet Sci.}  \bvol{12},  \pg{11--37}.

\bibitem[Huppert \& Turner(1981)]{huppert1981}
{\sc \au{Huppert, H.~E.} \& \au{Turner, J.~S.}} \yr{1981}  \at{Double-diffusive
  convection}.  \jt{J. Fluid Mech.}  \bvol{106},  \pg{299--329}.

\bibitem[Karniadakis {\em et~al.\/}(1991)Karniadakis, Israeli \&
  Orszag]{karniadakis1991high}
{\sc \au{Karniadakis, G.~E.}, \au{Israeli, M.} \& \au{Orszag, S.~A.}} \yr{1991}
   \at{High-order splitting methods for the incompressible {Navier}-{Stokes}
  equations}.  \jt{J. Comput. Phys.}  \bvol{97}~(2),  \pg{414--443}.

\bibitem[Kelley {\em et~al.\/}(2003)Kelley, Fernando, Gargett, Tanny \&
  \"{O}zsoy]{kelley2003}
{\sc \au{Kelley, D.~E.}, \au{Fernando, H. J.~S.}, \au{Gargett, A.~E.},
  \au{Tanny, J.} \& \au{\"{O}zsoy, E.}} \yr{2003}  \at{The diffusive regime of
  double diffusive convection}.  \jt{Prog. Oceanogr.}  \bvol{56},
  \pg{461--481}.

\bibitem[Knobloch(2015)]{knobloch2015}
{\sc \au{Knobloch, E.}} \yr{2015}  \at{Spatial localization in dissipative
  systems}.  \jt{Annu. Rev. Condens. Matter Phys.}  \bvol{6},  \pg{325--359}.

\bibitem[Knobloch {\em et~al.\/}(1986)Knobloch, Moore, Toomre \&
  Weiss.]{knobloch1986}
{\sc \au{Knobloch, E.}, \au{Moore, D.~R.}, \au{Toomre, J.} \& \au{Weiss.,
  N.~O.}} \yr{1986}  \at{Transitions to chaos in two-dimensional
  double-diffusive convection}.  \jt{J. Fluid Mech.}  \bvol{166},
  \pg{409--448}.

\bibitem[Knobloch {\em et~al.\/}(2019)Knobloch, Uecker \&
  Wetzel]{knobloch2019defectlike}
{\sc \au{Knobloch, E.}, \au{Uecker, H.} \& \au{Wetzel, D.}} \yr{2019}
  \at{Defectlike structures and localized patterns in the cubic-quintic-septic
  {Swift}\textendash{}{Hohenberg} equation}.  \jt{Phys. Rev. E}
  \bvol{100}~(1),  \pg{012204}.

\bibitem[Kolodner(1991)]{kolodner1991}
{\sc \au{Kolodner, P.}} \yr{1991}  \at{Stable and unstable pulses of
  traveling-wave convection}.  \jt{Phys. Rev. A}  \bvol{43},  \pg{2827--2832}.

\bibitem[Krishnamurti(2003)]{Krishnamurti2003}
{\sc \au{Krishnamurti, R.}} \yr{2003}  \at{Double-diffusive transport in
  laboratory thermohaline staircases}.  \jt{J. Fluid Mech.}  \bvol{483},
  \pg{287--314}.

\bibitem[Krishnamurti(2009)]{Krishnamurti2009}
{\sc \au{Krishnamurti, R.}} \yr{2009}  \at{Heat, salt and momentum transport in
  a laboratory thermohaline staircase}.  \jt{J. Fluid Mech.}  \bvol{638},
  \pg{491--506}.

\bibitem[Lay {\em et~al.\/}(2008)Lay, Hernlund \& Buffett]{Lay2008}
{\sc \au{Lay, T.}, \au{Hernlund, J.} \& \au{Buffett, B.~A.}} \yr{2008}
  \at{Core--mantle boundary heat flow}.  \jt{Nat. Geosci.}  \bvol{1},  \pg{25}.

\bibitem[Mamou {\em et~al.\/}(1998)Mamou, Vasseur \& Bilgen]{mamou1998double}
{\sc \au{Mamou, M.}, \au{Vasseur, P.} \& \au{Bilgen, E.}} \yr{1998}
  \at{Double-diffusive convection instability in a vertical porous enclosure}.
  \jt{J. Fluid Mech.}  \bvol{368},  \pg{263--289}.

\bibitem[Mamun \& Tuckerman(1995)]{mamun1995asymmetry}
{\sc \au{Mamun, C.~K.} \& \au{Tuckerman, L.~S.}} \yr{1995}  \at{Asymmetry and
  {Hopf} bifurcation in spherical {Couette} flow}.  \jt{Phys. Fluids}
  \bvol{7}~(1),  \pg{80--91}.

\bibitem[Matthews {\em et~al.\/}(1993)Matthews, Proctor, Rucklidge \&
  Weiss]{matthews1993pulsating}
{\sc \au{Matthews, P.~C.}, \au{Proctor, M. R.~E.}, \au{Rucklidge, A.~M.} \&
  \au{Weiss, N.~O.}} \yr{1993}  \at{Pulsating waves in nonlinear
  magnetoconvection}.  \jt{Phys. Lett. A}  \bvol{183}~(1),  \pg{69--75}.

\bibitem[Mercader {\em et~al.\/}(2009)Mercader, Batiste, Alonso \&
  Knobloch]{mercader2009}
{\sc \au{Mercader, I.}, \au{Batiste, O.}, \au{Alonso, A.} \& \au{Knobloch, E.}}
  \yr{2009}  \at{Localized pinning states in closed containers: homoclinic
  snaking without bistability}.  \jt{Phys. Rev. E}  \bvol{80},  \pg{025201}.

\bibitem[Mercader {\em et~al.\/}(2011)Mercader, Batiste, Alonso \&
  Knobloch]{mercader2011}
{\sc \au{Mercader, I.}, \au{Batiste, O.}, \au{Alonso, A.} \& \au{Knobloch, E.}}
  \yr{2011}  \at{Convectons, anticonvectons and multiconvectons in binary fluid
  convection}.  \jt{J. Fluid Mech.}  \bvol{667},  \pg{586--606}.

\bibitem[Paliwal \& Chen(1980{\natexlab{{\em a\/}}})]{paliwal1980a}
{\sc \au{Paliwal, R.~C.} \& \au{Chen, C.~F.}} \yr{1980{\natexlab{{\em a\/}}}}
  \at{Double-diffusive instability in an inclined fluid layer. part 1.
  experimental investigation}.  \jt{J. Fluid Mech.}  \bvol{98},  \pg{755--768}.

\bibitem[Paliwal \& Chen(1980{\natexlab{{\em b\/}}})]{paliwal1980b}
{\sc \au{Paliwal, R.~C.} \& \au{Chen, C.~F.}} \yr{1980{\natexlab{{\em b\/}}}}
  \at{Double-diffusive instability in an inclined fluid layer. part 2.
  stability analysis}.  \jt{J. Fluid Mech.}  \bvol{98},  \pg{769--785}.

\bibitem[P\'erez-Santos {\em et~al.\/}(2014)P\'erez-Santos, Garc\'es-Vargas,
  Schneider, Ross, Parra \& Valle-Levinson]{santos2014}
{\sc \au{P\'erez-Santos, I.}, \au{Garc\'es-Vargas, J.}, \au{Schneider, W.},
  \au{Ross, L.}, \au{Parra, S.} \& \au{Valle-Levinson, A.}} \yr{2014}
  \at{Double-diffusive layering and mixing in patagonian fjords}.  \jt{Prog.
  Oceanogr.}  \bvol{129},  \pg{35--49}.

\bibitem[Predtechensky {\em et~al.\/}(1994)Predtechensky, McCormich, Swift,
  Noszticzius \& Swinney]{Predtechensky1994}
{\sc \au{Predtechensky, A.~A.}, \au{McCormich, W.~D.}, \au{Swift, J.~B.},
  \au{Noszticzius, Z.} \& \au{Swinney, H.~L.}} \yr{1994}  \at{Onset of
  traveling waves in isothermal double diffusive convection}.  \jt{Phys. Rev.
  Lett.}  \bvol{72},  \pg{218--221}.

\bibitem[Requil{\'e} {\em et~al.\/}(2020)Requil{\'e}, Hirata, Ouarzazi \&
  Barletta]{requile2020weakly}
{\sc \au{Requil{\'e}, Y.}, \au{Hirata, S.~C.}, \au{Ouarzazi, M.~N.} \&
  \au{Barletta, A.}} \yr{2020}  \at{Weakly nonlinear analysis of viscous
  dissipation thermal instability in plane {Poiseuille} and plane {Couette}
  flows}.  \jt{J. Fluid Mech.}  \bvol{886}.

\bibitem[Rucklidge(1992)]{rucklidge1992}
{\sc \au{Rucklidge, A.~M.}} \yr{1992}  \at{Chaos in models of double
  convection}.  \jt{J. Fluid Mech.}  \bvol{237},  \pg{209--229}.

\bibitem[Schmitt(1983)]{schmitt1983characteristics}
{\sc \au{Schmitt, R.~W.}} \yr{1983}  \at{The characteristics of salt fingers in
  a variety of fluid systems, including stellar interiors, liquid metals,
  oceans, and magmas}.  \jt{Phys. Fluids}  \bvol{26}~(9),  \pg{2373--2377}.

\bibitem[Schmitt(1994)]{schmitt1994double}
{\sc \au{Schmitt, R.~W.}} \yr{1994}  \at{Double diffusion in oceanography}.
  \jt{Annu. Rev. Fluid Mech.}  \bvol{26},  \pg{255--285}.

\bibitem[Schmitt {\em et~al.\/}(2005)Schmitt, Ledwell, Montgomery, Polzin \&
  Toole]{schmitt2005enhanced}
{\sc \au{Schmitt, R.~W.}, \au{Ledwell, J.~R.}, \au{Montgomery, E.~T.},
  \au{Polzin, K.~L.} \& \au{Toole, J.~M.}} \yr{2005}  \at{Enhanced diapycnal
  mixing by salt fingers in the thermocline of the tropical {Atlantic}}.
  \jt{Science}  \bvol{308}~(5722),  \pg{685--688}.

\bibitem[Schmitt {\em et~al.\/}(1987)Schmitt, Perkins, Boyd \&
  Stalcup]{schmitt1987csalt}
{\sc \au{Schmitt, R.~W.}, \au{Perkins, H.}, \au{Boyd, J.~D.} \& \au{Stalcup,
  M.~C.}} \yr{1987}  \at{{C-SALT: An} investigation of the thermohaline
  staircase in the western tropical {North Atlantic}}.  \jt{Deep-Sea Res.}
  \bvol{34}~(10),  \pg{1655--1665}.

\bibitem[Shankar {\em et~al.\/}(2021)Shankar, Kumar \&
  Shivakumara]{shankar2021stability}
{\sc \au{Shankar, B.~M.}, \au{Kumar, J.} \& \au{Shivakumara, I.~S.}} \yr{2021}
  \at{Stability of double-diffusive natural convection in a vertical fluid
  layer}.  \jt{Phys. Fluids}  \bvol{33}~(9),  \pg{094113}.

\bibitem[Spiegel(1969)]{spiegel1969}
{\sc \au{Spiegel, E.~A.}} \yr{1969}  \at{Semiconvection}.  \jt{Comments on
  Astrophysics and Space Physics}  \pg{pp. 1--57}.

\bibitem[Spiegel(1972)]{spiegel1972}
{\sc \au{Spiegel, E.~A.}} \yr{1972}  \at{Convection in stars ii. special
  effects}.  \jt{Annu. Rev. of Astron. Astrophys.}  \bvol{10},  \pg{261--304}.

\bibitem[Spina {\em et~al.\/}(1998)Spina, Toomre \& Knobloch]{spina1998}
{\sc \au{Spina, A.}, \au{Toomre, J.} \& \au{Knobloch, E.}} \yr{1998}
  \at{Confined states in large-aspect-ratio thermosolutal convection}.
  \jt{Phys. Rev. E}  \bvol{57},  \pg{524--547}.

\bibitem[Tsitverblit(1995)]{tsitverblit1995bifurcation}
{\sc \au{Tsitverblit, N.}} \yr{1995}  \at{Bifurcation phenomena in confined
  thermosolutal convection with lateral heating: Commencement of the
  double-diffusive region}.  \jt{Phys. Fluids}  \bvol{7}~(4),  \pg{718--736}.

\bibitem[Tsitverblit \& Kit(1993)]{tsitverblit1993multiplicity}
{\sc \au{Tsitverblit, N.} \& \au{Kit, E.}} \yr{1993}  \at{The multiplicity of
  steady flows in confined double-diffusive convection with lateral heating}.
  \jt{Phys. Fluids A}  \bvol{5}~(4),  \pg{1062--1064}.

\bibitem[Umbr{\'\i}a \& Net(2019)]{umbria2019stationary}
{\sc \au{Umbr{\'\i}a, J.~S.} \& \au{Net, M.}} \yr{2019}  \at{Stationary {Flows}
  and {Periodic} {Dynamics} of {Binary} {Mixtures} in {Tall} {Laterally}
  {Heated} {Slots}}.  \bt{In {\em Computational Modelling of Bifurcations and
  Instabilities in Fluid Dynamics\/}},  \pg{pp. 171--216}.  \publ{Springer}.

\bibitem[Watanabe {\em et~al.\/}(2012)Watanabe, Iima \& Nishiura]{watanabe2012}
{\sc \au{Watanabe, T.}, \au{Iima, M.} \& \au{Nishiura, Y.}} \yr{2012}
  \at{Spontaneous formation of travelling localized structures and their
  asymptotic behaviour in binary fluid convection}.  \jt{J. Fluid Mech.}
  \bvol{712},  \pg{219--243}.

\bibitem[Watanabe {\em et~al.\/}(2016)Watanabe, Iima \& Nishiura]{watanabe2016}
{\sc \au{Watanabe, T.}, \au{Iima, M.} \& \au{Nishiura, Y.}} \yr{2016}  \at{A
  skeleton of collision dynamics: hierarchical network structure among
  even-symmetric steady pulses in binary fluid convection}.  \jt{SIAM J. Appl.
  Dyn. Syst.}  \bvol{15},  \pg{789--806}.

\bibitem[Wilcox(1993)]{wilcox1993}
{\sc \au{Wilcox, W.~R.}} \yr{1993}  \at{Transport phenomena in crystal growth
  from solution}.  \jt{Prog. Cryst. Growth Charact. Mater.}  \bvol{26},
  \pg{153--194}.

\bibitem[Xin {\em et~al.\/}(1998)Xin, Le~Qu{\'e}r{\'e} \&
  Tuckerman]{xin1998bifurcation}
{\sc \au{Xin, S.}, \au{Le~Qu{\'e}r{\'e}, P.} \& \au{Tuckerman, L.~S.}}
  \yr{1998}  \at{Bifurcation analysis of double-diffusive convection with
  opposing horizontal thermal and solutal gradients}.  \jt{Phys. Fluids}
  \bvol{10}~(4),  \pg{850--858}.

\bibitem[You(2002)]{you2002}
{\sc \au{You, Y.}} \yr{2002}  \at{A global ocean climatological atlas of the
  turner angle: Implications for double-diffusion and water-mass structure}.
  \jt{Deep-Sea Res.}  \bvol{49},  \pg{2075--2093}.

\end{thebibliography}
	
\end{document}